\documentclass[%
 reprint,
superscriptaddress,
nofootinbib,
 amsmath,amssymb,
 aps,
]{revtex4-1}

\usepackage{graphicx}% Include figure files
\usepackage{dcolumn}% Align table columns on decimal point
\usepackage{bm}% bold math
\usepackage{braket}
\usepackage{upgreek}
\usepackage{xcolor}
\usepackage{afterpage}
\usepackage{relsize}
\usepackage{multirow}
\usepackage{float}
\usepackage{makecell}
\usepackage{placeins}
\usepackage[colorlinks=true,urlcolor=blueprl,citecolor=blueprl,linkcolor=blueprl]{hyperref}
\usepackage{mathtools} % For \mathmakebox
\usepackage{changepage} % For adjustwidth
\usepackage{lineno}
\newcommand{\ketbra}[2]{\ket{#1}\!\bra{#2}}

\newcommand{\ketbraauto}[1]{\ketbra{#1}{#1}}

\newcolumntype{P}[1]{>{\centering\arraybackslash}p{#1}}
\newcommand{\ZJ}[1]{\textcolor{black}{#1}}
\newcommand{\ozlem}[1]{\textcolor{black}{#1}}
\newcommand{\ozlemREV}[1]{\textcolor{black}{#1}}

\definecolor{blueprl}{RGB}{46,48,146}

\def\ANU{Centre of Excellence for Quantum Computation and Communication Technology, The Department of Quantum Science and Technology, Research School of Physics and Engineering, The Australian National University, Canberra, Australian Capital Territory, Australia}
\def\ASTAR{A*STAR Quantum Innovation Centre (Q.InC), Agency~for~Science,~Technology~and~Research~(A*STAR), 2 Fusionopolis Way, Innovis \#08-03, Singapore 138634, Republic of Singapore}
\def\UQ{Centre for Quantum Computation and Communication Technology, School of Mathematics and Physics, University of Queensland, St Lucia, QLD 4072, Australia}
\def\CSIRO{Data61, Commonwealth Scientific and Industrial Research Organisation, Marsfield 2122, NSW, Australia}

\begin{document}

%\preprint{APS/123-QED}

\title{Capacity-Achieving Entanglement Purification Protocol for Pauli Dephasing Channel}

\author{\"{O}zlem Erk{\i}l{\i}\c{c}}
\email{ozlemerkilic1995@gmail.com}
\affiliation{\ANU}
\author{Matthew S. Winnel}
\affiliation{\UQ}
\author{Aritra Das}
\affiliation{\ANU}
\author{Sebastian Kish}
\affiliation{\CSIRO}
\author{Ping Koy Lam}
\affiliation{\ASTAR}
\affiliation{\ANU}
%\affiliation{\ASTARR}
\author{Jie Zhao}
\affiliation{\ANU}
\author{Syed M. Assad}
\email{cqtsma@gmail.com}
\affiliation{\ASTAR}
\affiliation{\ANU}
%\affiliation{\ASTARR}

\date{\today}

%\tableofcontents

\begin{abstract}
Quantum communication enables secure information transmission and entanglement distribution, but these tasks are fundamentally limited by the capacities of quantum channels. While quantum repeaters can mitigate losses and noise, entanglement swapping via a central node is ineffective against the Pauli dephasing channel due to degradation from Bell-state measurements. This suggests that purifying distributed Bell states before entanglement swapping is necessary. \ozlemREV{Although one-way hashing codes are known to saturate the dephasing channel capacity, no explicit two-way purification protocol has previously been shown to achieve this bound.} In this work, we present a two-way entanglement purification protocol with an explicit, scalable circuit that asymptotically achieves the dephasing channel capacity. With each iteration, the fidelity of Bell states increases.
%producing perfect Bell pairs in the limit of many iterations, making them ideal for use in quantum repeaters and for correcting dephasing errors in quantum computers. 
\ozlemREV{At the final round, the residual dephasing error is suppressed doubly-exponentially, scaling as $\mathcal{O}(p^{2^{n}})$, enabling near-perfect Bell pairs for any fixed number of purification rounds $n$.} The explicit circuit we propose is versatile and applicable to any number of Bell pairs, offering a practical solution for mitigating decoherence in quantum networks and distributed quantum computing.
\end{abstract}

\maketitle

\section{\label{sec:level1}Introduction}
Quantum communications~\cite{gisin2007quantum} guarantee the reliable transmission of quantum information, efficient distribution of entanglement~\cite{dynes2009efficient, inagaki2013entanglement}, and enable the establishment of secure secret keys~\cite{pirandola2020advances, xu2020secure}. However, the achievable rates for these tasks are inherently constrained by the capacity of the quantum channels~\cite{pirandola2009direct, wilde2017converse, pirandola2017fundamental, pirandola2019end}. These quantum channels often cause some decoherence on the distributed quantum states resulting in a degradation of the quality of the shared entanglement or the rates at which secret keys can be generated. Designing optimal protocols in the presence of inevitable decoherence is necessary for the realisation of quantum networks~\cite{kimble2008quantum}.

Pauli dephasing channel is an example of such a channel where the main source of decoherence is a phase flip~\cite{nielsen2010quantum}. The amount of distributed entanglement in a given channel is known as the entanglement flux which gives the upper bound of the channel capacity~\cite{pirandola2017fundamental}, whereas the reverse coherent information (RCI)~\cite{pirandola2009direct, garcia2009reverse} is a lower bound that measures the amount of the distillable entanglement of a quantum state, which is achievable by an optimal protocol based on one-way classical communication. As these two quantities coincide with each other for the dephasing channel, this channel is distillable~\cite{pirandola2017fundamental}. The capacity of this channel assisted by two-way classical communications for point-to-point communication (i.e., when two parties are directly communicating with each other) is given by $1-\mathrm{H}_2(p)$ where $\mathrm{H}_2$ is the binary Shannon entropy and $p\in[0,0.5]$ is the dephasing probability of the channel. This bound is an achievable rate for the key rate generation, two-way entanglement distribution and two-way quantum capacity~\cite{pirandola2017fundamental}.

Similar to the capacity of the point-to-point communications, also known as the repeaterless bound, there are also fundamental limits for the achievable rates for end-to-end communications~\cite{pirandola2019end}, that incorporate quantum repeaters. Quantum repeaters are devices that divide the quantum channel into smaller segments to make the decoherence of the channel more manageable~\cite{briegel1998quantum, dur1999quantum, munro2015inside, azuma2022quantum}. Quantum repeaters use entanglement swapping~\cite{goebel2008multistage,kaltenbaek2009high} to distribute entanglement between the trusted parties. The point-to-point capacity of the dephasing channel can be saturated with Alice distributing maximally entangled Bell states to Bob and to achieve the highest secret key rate generation, Bob needs to perform a measurement in the Hadamard basis. The use of a quantum repeater is necessary to surpass the repeaterless bound~\cite{munro2015inside,muralidharan2016optimal}. In other quantum channels such as the pure-loss channel where the only source of decoherence is the photon losses, the use of a repeater can exceed the point-to-point bound~\cite{lucamarini2018overcoming, wang2018twin, cui2019twin, dias2020quantum, winnel2021overcoming, erkilicc2023surpassing}. However, having a node in the dephasing channel does not yield better key rates than the direct communication between Alice and Bob. For instance, consider a scenario where Alice and Bob send maximally entangled Bell states to a quantum node. The node performs a Bell-state measurement to establish an entangled state between Alice and Bob. Interestingly, the RCI of this state remains the same as it would be without using a node, despite the lower effective decoherence before the entanglement swapping measurement. 
%For example, let us consider a scenario where Alice and Bob send maximally entangled Bell states to a repeater node where the repeater uses a Bell-state measurement to establish an entangled state between Alice and Bob. The RCI of this state coincides with the RCI of the state without a repeater, even though the effective decoherence before the entanglement swapping measurement is less in comparison to the direct transmission. 
This indicates that the state needs to be purified before the node can perform an entanglement swapping measurement using entanglement purification or error correction.

Entanglement purification~\cite{bennett1996purification, bennett1996mixed, deutsch1996quantum, horodecki1997inseparable, sheng2010deterministic, sheng2010one, sheng2013hybrid, hu2021long, winnel2022achieving, huang2022experimental} is used to concentrate the total entanglement of a large number of noisy entangled states into a fewer number of states with higher purity using local operations and measurements by the parties sharing the states. Entanglement purification protocols can be classified into filtering, recurrence, hashing and breeding protocols~\cite{dur2007entanglement}. Filtering protocols~\cite{verstraete2001local} utilise a single copy of a mixed state and apply some local filtering measurements such as weak measurements to yield a state with better entanglement. However, the achievable fidelity of the output state is limited by the probability of success and these protocols have limited capabilities for general mixed states~\cite{dur2007entanglement}. In contrast, recurrence protocols~\cite{bennett1996purification, bennett1996mixed, deutsch1996quantum, dur2003entanglement} operate on two copies of mixed states at a time by performing some local operations on both states and measuring one. This procedure is repeated until the desired fidelity is achieved. Even though these protocols can produce a maximally entangled pure state given enough iterations, the rates of these protocols cannot saturate the ultimate two-way capacities due to measuring half of the states in each purification step. Unlike recurrence protocols, $N\rightarrow M$ protocols~\cite{dehaene2003local} use many copies of mixed states at a time where $N-M$ states out of $N$ states are measured. Hashing protocols~\cite{bennett1996mixed} are a special case of $N\rightarrow M$ protocols where they operate on a \ZJ{large} number of mixed states. In hashing protocols, both parties sharing the entangled states apply controlled-not, CNOT, gates and other local unitary operations on their own pairs and measure some of the entangled states to get information about the remaining entangled pairs. Breeding protocols work in a similar fashion, however, they also use pre-purified entangled states that are utilised to gain information about the unmeasured states. It is known that the hashing protocols assisted with one-way classical communication~\cite{bennett1996mixed, devetak2005distillation} can achieve the distillable entanglement capacity. 
%However, the optimal protocol that can reach this bound using entanglement purification assisted with two-way classical communication is unknown~\cite{dur2007entanglement}.
\ozlem{While one-way communication is a subset of two-way classical communication, no protocol leveraging the additional features of two-way communication has been shown to reach this bound~\cite{dur2007entanglement}.}

Quantum error correction (QEC) is crucial for purifying entangled states in noisy quantum channels by detecting and correcting errors without disturbing the quantum states~\cite{devitt2013quantum}. Various QEC codes exist, such as repetition codes~\cite{wootton2018repetition}, quantum low-density parity-check (QLDPC) codes~\cite{postol2001proposed, mackay2004sparse, kasai2011quantum}, and stabiliser codes~\cite{shor1995scheme, steane1996error, calderbank1996good, gottesman1997stabilizer, knill2001benchmarking}. Repetition codes are simple but less efficient, mainly correcting bit-flip errors. QLDPC codes offer efficient error correction with high tolerance using a sparse parity-check matrix. Stabiliser codes, like Shor's~\cite{shor1995scheme}, Steane's~\cite{steane1996error} and CSS codes~\cite{calderbank1996good}, encode quantum information into a larger Hilbert space, using a stabiliser group to detect and correct various errors. QEC has been shown to be equivalent to entanglement purification protocols using one-way classical communication~\cite{bennett1996mixed, dur2007entanglement}. Specifically, CSS codes for Pauli-diagonal channels can be adapted for entanglement purification with one-way or two-way communication~\cite{matsumoto2003conversion, aschauer2005quantum, ambainis2006minimum}. Two-way protocols are particularly effective as they allow for selectively discarding certain entangled pairs, resulting in higher fidelity. In contrast, one-way protocols do not permit discarding pairs without losing information, making two-way purification protocols derived from CSS codes more effective~\cite{dur2007entanglement}.

While it is known that the Pauli dephasing channel can be saturated using entanglement purification codes assisted by one-way classical communication, as it is a distillable channel, \ozlemREV{no explicit protocol leveraging two-way classical communication has previously been shown to achieve this capacity.} In this work, \ozlemREV{we introduce the first two-way entanglement-purification protocol with a fully explicit and physically implementable circuit that asymptotically attains the dephasing-channel capacity.}
%In this work, we propose a two-way entanglement purification protocol with an explicit circuit that achieves the dephasing channel capacity in the asymptotic limit.
Alice prepares $m$ pairs and sends them to Bob. Both parties then apply CNOT gates locally on their dephased qubits, with the $m^{th}$ qubit being the control and the others being the targets. After applying the CNOT gates, the control qubit is measured in the Hadamard basis. This recursive protocol involves Alice and Bob purifying the states in multiple stages, resulting in the remaining Bell pairs becoming purer with each iteration. By the final round of the protocol, the Bell pairs will have a fidelity close to one with respect to $\ket{\phi^+}=(\ket{00}+\ket{11})/\sqrt{2}$, making them suitable for use in a repeater protocol to distribute entanglement at the channel capacity. Although achieving capacity requires asymptotically large $m$, \ozlemREV{the protocol remains effective for finite $m$ and a finite number of rounds $n$,} making it practical for realistic quantum-network settings. \ozlemREV{In particular, each purification round suppresses the dephasing probability quadratically, and by the $n^{\text{th}}$ round the residual error decreases doubly-exponentially, scaling as $\mathcal{O}(p^{2^{n}})$.} \ozlemREV{Consequently, only $n = \mathcal{O}\!\big(\log(|\log\delta|)\big)$ rounds are required to reach a target fidelity of $1-\delta$, while the associated reduction in yield remains mild, with $Y = \big((m-1)/m\big)^{n}$.} Beyond repeater applications, the same circuit can also be used to purify Bell pairs affected by dephasing noise in quantum computing settings. Furthermore, this protocol can serve as a foundational framework for addressing noise and mitigating decoherence in non-distillable channels, such as thermal-noise and depolarising channels, where two-way classical communication could be more advantageous, but would need to be tailored to suit the specific characteristics of those channels.

The paper is structured as follows. In section~\ref{sec:details_of_protocol}, we introduce the purification protocol and explain the circuit design for arbitrary number of Bell pairs, followed by the main analytical results and numerical simulations, including the scaling of the RCI, the protocol’s convergence to capacity and the fidelity of the purified states. In section~\ref{sec:conclusions}, we summarise our results and conclude with an outlook.

\section{\label{sec:details_of_protocol}Details of the Protocol}
\begin{figure}[h!]
%\hspace*{-0.3cm}
\includegraphics[scale=0.2]{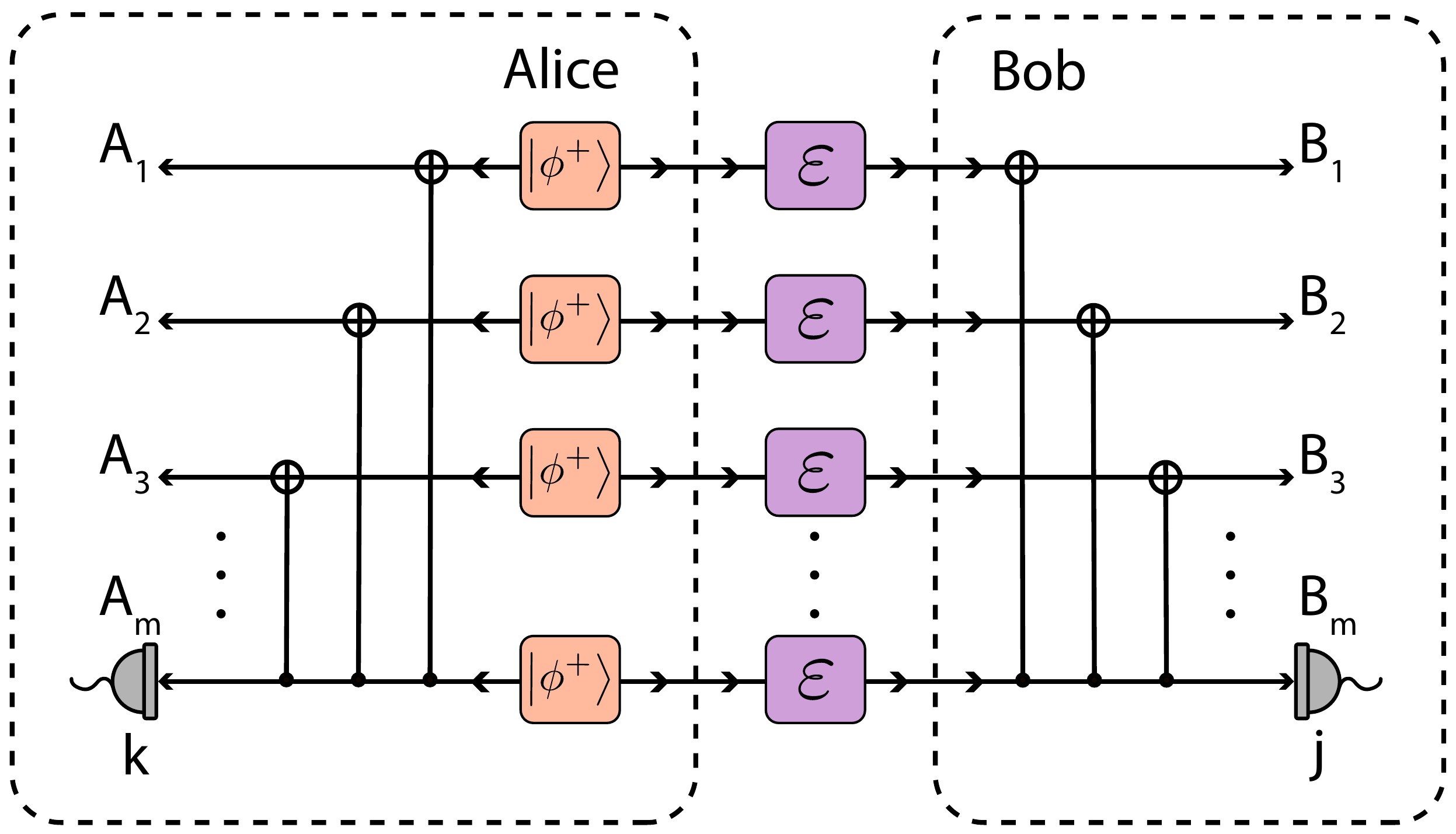}
\caption{\label{fig:stage1} First stage of the protocol: Alice prepares $m$ copies of $\ket{\phi^+}$ Bell states and sends them to Bob via a Pauli dephasing channel. Both apply CNOT gates on their qubits, with $\mathrm{A_1, A_2}, \ldots, \mathrm{A_m}$ and $\mathrm{B_1, B_2}, \ldots, \mathrm{B_m}$ representing their respective modes. After the CNOTs, $\mathrm{A_m}$ and $\mathrm{B_m}$ are measured in the Hadamard basis.}
\end{figure}
Assume we have a maximally entangled Bell state $\ket{\phi^+}$ in the form
\begin{equation}
    \label{eq:phi_plus}
    \ket{\phi^+}=\frac{\ket{00}+\ket{11}}{\sqrt{2}}.
\end{equation}
In the dephasing channel without a repeater, this pure state becomes a mixture of $\ket{\phi^+}$ and $\ket{\phi^-}$ given by
\begin{equation}
    \rho_{\mathrm{AB}}=(1-p)\ketbraauto{\phi^{+}}+p\ketbraauto{\phi^{-}},
    \label{eq:choi}
\end{equation}
where $\ket{\phi^{-}}=(\ket{00}-\ket{11})/ \sqrt{2}$ and $p$ is the dephasing probability, also known as the probability of a phase-flip with $0\leq p\leq1/2$. Note that when $p=1/2$, the state becomes completely dephased. If there is a \ZJ{node in the mid-point} splitting the channel into two, the effective state between the sender (Alice) and the \ozlem{node (Charlie)} is given by
\begin{multline}
    \label{eq:choi_repeater}
    \rho_{\text{AC}}=\bigg(1-\frac{1-\sqrt{1-2p}}{2}\bigg)\ketbraauto{\phi^{+}}\\+\bigg(\frac{1-\sqrt{1-2p}}{2}\bigg)\ketbraauto{\phi^{-}}.
\end{multline}

Even though the state dephases less due to having a node, when Bell-state measurements (BSM) are used to swap entanglement, the resulting state between Alice and Bob is the same as equation~\eqref{eq:choi} (\ozlem{refer to appendix~\ref{sec:appendix_repeater_proof} for the details).} This indicates the necessity of purifying state before performing entanglement swapping.

\subsection{\label{sec:first_stage}The First Stage of the Protocol}
In this protocol, Alice prepares $m$ copies of Bell pairs and distributes them to Bob via a quantum channel as shown in figure~\ref{fig:stage1}. One half of each Bell pair undergoes dephasing with a probability of $p$ during transmission. To mitigate the dephasing effects and enhance the purity of the Bell pairs, Alice and Bob apply CNOT gates \ZJ{between their respective $m^{\mathrm{th}}$ qubit and all other qubits.} They then measure the $m^\mathrm{th}$ qubits, where the measured qubit acts as the control and the remaining qubits serve as the targets. Alice and Bob's measurement is performed using the eigenvectors of the $X$ basis given as
\begin{equation}
    \label{eq:eigenvectorsX}
    \ket{\pm}=\frac{\ket{0}\pm\ket{1}}{\sqrt{2}}.
\end{equation}
When Alice's and Bob's measurement outcomes, denoted as $j$ and $k$ respectively, match as either $(++)$ or $(--)$, they successfully distill entanglement, resulting in a slightly purified state. However, if the outcomes differ, resulting in $(+-)$ or $(-+)$, the distillation process fails. To determine this, Alice and Bob exchange their measurement results via two-way classical communication. This step ensures that both parties agree on whether the purification was successful or failed. As previously mentioned, this protocol is a recursive purification process. Regardless of whether the distillation is successful, Alice and Bob move on to the next stage to further purify the Bell pairs, as explained in the following section.

In the first stage, Alice and Bob apply CNOT gates, leaving the target qubits unchanged. The measured Bell pair (the control qubit) switches to $\ket{\phi^-}$ if there are an odd number of phase flips, and to $\ket{\phi^+}$ if there are an even number. For example, with three Bell pairs, as shown in Table~\ref{tab:table_first_stage}, an even number of bit-flips results in a $\ket{\phi^+}$ state. A successful outcome is indicated when Alice and Bob both measure $(++)$ or $(--)$, ensuring the measured Bell pair is always $\ket{\phi^+}$. This measurement selects states with $0, 2, \cdots$ bit flips thus removing first-order dephasing errors. The unnormalised eigenvalues of the joint state of the remaining pairs are then based on the probabilities of even bit-flips,
\begin{equation}
    \Tilde{\lambda}_{s_1}(j)=(1-p)^{m-j}p^j,
\label{eq:eigenvalues}
\end{equation}
where $j$ represents the number transmission errors with $j=0, 2, \ldots, m$ and each eigenvalue has a multiplicity of $\binom{m}{j}$. \ZJ{The normalised eigenvalues are given by}
\begin{equation}
    \lambda_{s_1}(j)=\frac{\Tilde{\lambda}_{s_1}(j)}{P_{s_1}(m)},
\label{eq:normedeigenvalues_stage_1}
\end{equation}
\ZJ{where the normalisation factor}
%It is important to normalise each eigenvalue, with the normalisation factor, also known as the probability of success for this round, given by
\begin{equation}
    P_{\mathrm{s}_1}(m)=\frac12 \left (1+(1-2p)^m \right ),
\label{eq:prob_success_stage1}
\end{equation}
\ZJ{is the probability of success for this round.}

\begin{table}[t]
\caption{\label{tab:table_first_stage}
Truth table of the first stage of the protocol where Alice distributes three Bell pairs. T and C denote target and control Bell pairs, respectively, with the control being measured. \ZJ{The first and second columns show the state before and after applying the CNOT gates, where only the control qubit is affected. The third column presents the probability of dephasing noise occurring after distributing three Bell pairs.} Pink entries indicate the accepted outcomes when Alice and Bob get a matching measurement outcome.
}
\begin{ruledtabular}
\begin{tabular}{ccc}
\textrm{T,T,C}&
\textrm{T,T,C}&
\textrm{Probability}\\
\colrule
 $\phi^+, \phi^+, \phi^+$  & \textcolor{magenta}{$\phi^+, \phi^+, \phi^+$} & $(1-p)^3$ \\
 $\phi^+, \phi^+, \phi^-$ & $\phi^+, \phi^+, \phi^-$ & $p(1-p)^2$ \\
 $\phi^+, \phi^-, \phi^+$ & $\phi^+, \phi^-, \phi^-$ & $p(1-p)^2$ \\
 $\phi^+, \phi^-, \phi^-$ & \textcolor{magenta}{$\phi^+, \phi^-, \phi^+$} & $p^2(1-p)$ \\
  $\phi^-, \phi^+, \phi^+$ & $\phi^-, \phi^+, \phi^-$ & $p(1-p)^2$ \\
$\phi^-, \phi^+, \phi^-$ & \textcolor{magenta}{$\phi^-, \phi^+, \phi^+$} & $p^2(1-p)$ \\
 $\phi^-, \phi^-, \phi^+$ & \textcolor{magenta}{$\phi^-, \phi^-, \phi^+$} & $p^2(1-p)$ \\
  $\phi^-, \phi^-, \phi^-$ & $\phi^-, \phi^-, \phi^-$ & $p^3$
\end{tabular}
\end{ruledtabular}
\end{table}

For the mismatched outcomes $(+-)$ or $(-+)$, the eigenvalues of the conditional state are derived from an odd number of phase-flips, calculated using equation~\eqref{eq:eigenvalues} with $j=1, 3, \ldots, m$ and denoted as $\lambda_{f_1}(j)$. Each eigenvalue has a multiplicity of $\binom{m}{j}$. The probability of failure is given by
\begin{equation}
    \label{eq:prob_failure_stage1}
    P_{f_1}(m)=1-P_{s_1}(m)=\frac{1}{2}(1-(1-2p)^m).
\end{equation}

This protocol maps individually distributed Bell pairs to a higher-dimensional state. For instance, if Alice and Bob share $m=3$ dephased Bell pairs, after the first round, the remaining state between them becomes a diagonal state in the Bell basis, as shown below
\begin{align}
 \rho_{\mathrm{A_1B_1A_2B_2}}=&\lambda_{s_1}(0)\ketbraauto{\phi^+\phi^+}\nonumber \\
 &+\lambda_{s_1}(1)\ketbraauto{\phi^+\phi^-}\nonumber \\
 &+\lambda_{s_1}(1)\ketbraauto{\phi^-\phi^+}\nonumber\\
 &+\lambda_{s_1}(2)\ketbraauto{\phi^-\phi^-}.
\label{eq:example_state}   
\end{align}
Here, $\lambda_{s_1}(0)$, $\lambda_{s_1}(1)$, and $\lambda_{s_1}(2)$ are the eigenvalues of this state. As seen in equation~\eqref{eq:example_state}, the remaining state now consists of \ZJ{a mixture of} 2 Bell pairs between Alice and Bob, with each eigenvalue indicating the number of errors left after measuring the third Bell pair. Specifically, $\lambda_{s_1}(0)$ corresponds to no errors, while $\lambda_{s_1}(1)$ and $\lambda_{s_1}(2)$ represent 1 and 2 errors, respectively.

After the first round of the protocol, we can calculate the RCI. The RCI is used to provide a lower bound on the distillable entanglement of a given channel~\cite{garcia2009reverse} and measures the transmission of quantum information. Since the Pauli dephasing channel is distillable, RCI serves as a tight bound, indicating the amount of distillable entanglement. To saturate the capacity of the dephasing channel, we also need to keep the unsuccessful states (i.e., states corresponding to mismatched measurement outcomes), as these states still contain useful entanglement between Alice and Bob, consistent with findings in~\cite{zhou2020purification}. Unlike conventional entanglement purification protocols that focus on reaching a target fidelity with minimal rounds or input resources, our objective here is to develop a protocol that asymptotically achieves the dephasing channel capacity. As this capacity is defined by the RCI and is fundamentally difficult to saturate, requiring asymptotically large numbers of Bell pairs and recursive operations, we adopt the RCI of the output state as our primary figure of merit. Therefore, after the first stage of the protocol, the RCI between Alice and Bob can be expressed as
\begin{figure}[h!]
%\hspace*{-0.3cm}
\includegraphics[scale=0.53]{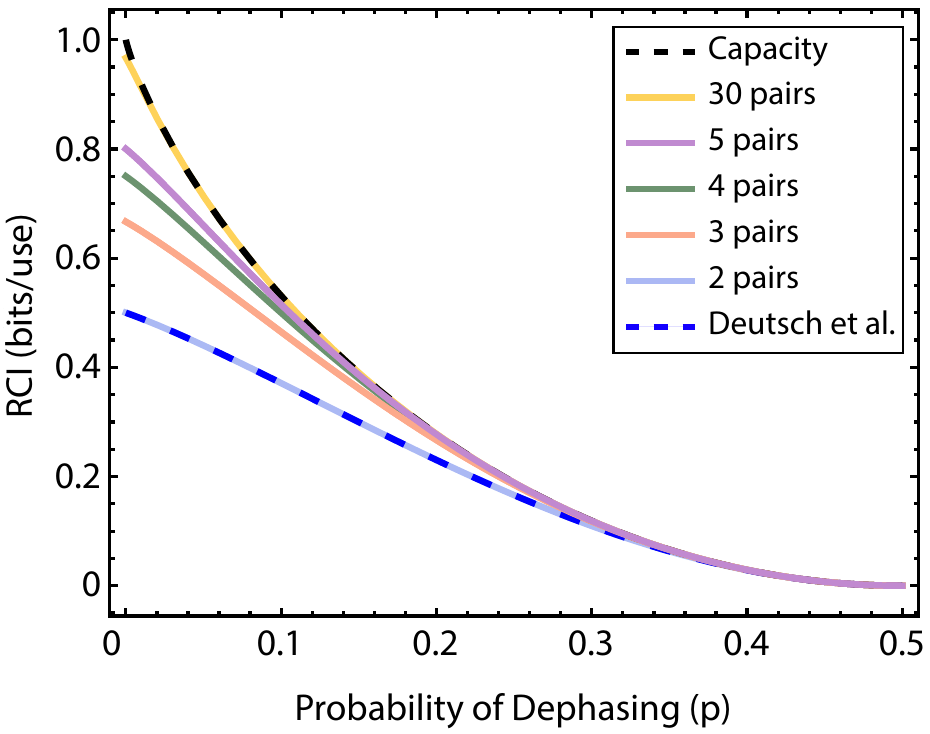}
\caption{\label{fig:stage1_result} Simulation results of the first round of the protocol with increasing number of Bell pairs $m$ shared between Alice and Bob. The black dashed line represents the dephasing channel capacity, while the blue dashed line shows results from Deutsch \textit{et al.}~\cite{deutsch1996quantum}.}
\end{figure}
\begin{equation}
    \mathrm{RCI}_1=\frac{1}{m}\big(P_{s_1}(m)\mathrm{RCI}(\rho_{\mathrm{AB}_{s_1}})+P_{f_1}(m)\mathrm{RCI}(\rho_{\mathrm{AB}_{f_1}})\big),
    \label{eq:average_rci_stage1}
\end{equation}
where $P_{s_1}(m)$ and $P_{f_1}(m)$ represent the probabilities of success and failure of Alice and Bob's measurements, respectively, while $\rho_{\mathrm{AB}_{s_1}}$ and $\rho_{\mathrm{AB}_{f_1}}$ are the joint states between Alice and Bob corresponding to successful and failed measurements. Note that the overall RCI needs to be divided by $m$, the number of Bell pairs used, as the channel capacity is expressed as per channel use. This normalisation by $m$, together with the inclusion of both successful and failed states weighted by their respective probabilities, ensures that the RCI expressions we derive (e.g., equations~\eqref{eq:average_rci_stage1}, \eqref{eq:rci_one_full} and~\eqref{eq:average_rci_stage2}) reflect both the probabilistic nature of the protocol and the total resource cost. As such, the RCI serves as an effective entanglement rate per channel use, analogous to the conventional output-to-input ratio of purified Bell states required to reach a target fidelity.

The RCI is computed using the von Neumann entropy as follows
\begin{equation}
    \mathrm{RCI}(\rho_{\mathrm{AB}})=S(\rho_\mathrm{A})-S(\rho_{\mathrm{AB}}),
    \label{eq:rci_equation}
\end{equation}
where $S(\rho_\mathrm{A})$ represents the von Neumann entropy of Alice's qubits, which is always equal to 1 when Alice and Bob share a single Bell pair. However, when they share $m$ Bell pairs and measure one out of the $m$ pairs, $S(\rho_\mathrm{A})=\log_2(2^{m-1})=m-1$. $S(\rho_{\mathrm{AB}})$ represents the entropy of Alice and Bob's joint state which is calculated from 
\begin{subequations}
\label{eq:entropy_joint_stage1}
\begin{equation}
    \label{eq:entropy_joint_stage1_success}
    S(\rho_{\mathrm{AB_{s_1}}})=-\sum_{j=0,2,\cdots}^{m}\binom{m}{j}\lambda_{s_1}(j)\mathrm{log_2}\lambda_{s_1}(j),
\end{equation}
\begin{equation}
    \label{eq:entropy_joint_stage1_fail}
    S(\rho_{\mathrm{AB}_{f_1}})=-\sum_{j=1,3,\cdots}^{m}\binom{m}{j}\lambda_{f_1}(j)\mathrm{log_2}\lambda_{f_1}(j),  
\end{equation}
\end{subequations}
where $\binom{m}{j}$ denotes the multiplicity of each eigenvalue $\lambda_{s_1}(j)$ and  $\lambda_{f_1}(j)$. equation~\eqref{eq:entropy_joint_stage1_success} presents the joint entropy of the successful state, while equation~\eqref{eq:entropy_joint_stage1_fail} provides the entropy of the failed state. Therefore, the overall RCI of the first stage becomes
\begin{align}
    \label{eq:rci_one_full}
    \mathrm{RCI_1}=&\frac{1}{m}\bigg[(m-1)+\!P_{s_1}(m)\hspace{-0.5em}\!\sum_{j=0,2,\cdots}^{m}\hspace{-0.5em}\!\binom{m}{j}\lambda_{s_1}(j)\mathrm{log_2}\lambda_{s_1}(j)\nonumber\\
    &+P_{f_1}(m)\hspace{-0.5em}\!\sum_{j=1,3,\cdots}^{m}\hspace{-0.5em}\!\binom{m}{j}\lambda_{f_1}(j)\mathrm{log_2}\lambda_{f_1}(j)\bigg].
\end{align}
Note that the term $(m-1)$ comes from $P_{s_1}(m)S(\rho_{\mathrm{A}_{s_1}})+P_{f_1}(m)S(\rho_{\mathrm{A}_{f_1}})$ where $S(\rho_{\mathrm{A}_{s_1}})=S(\rho_{\mathrm{A}_{f_1}})=m-1$ and $P_{s_1}(m)+P_{f_1}(m)=1$.
\begin{figure}[t!]
%\hspace*{-0.3cm}
\includegraphics[scale=0.19]{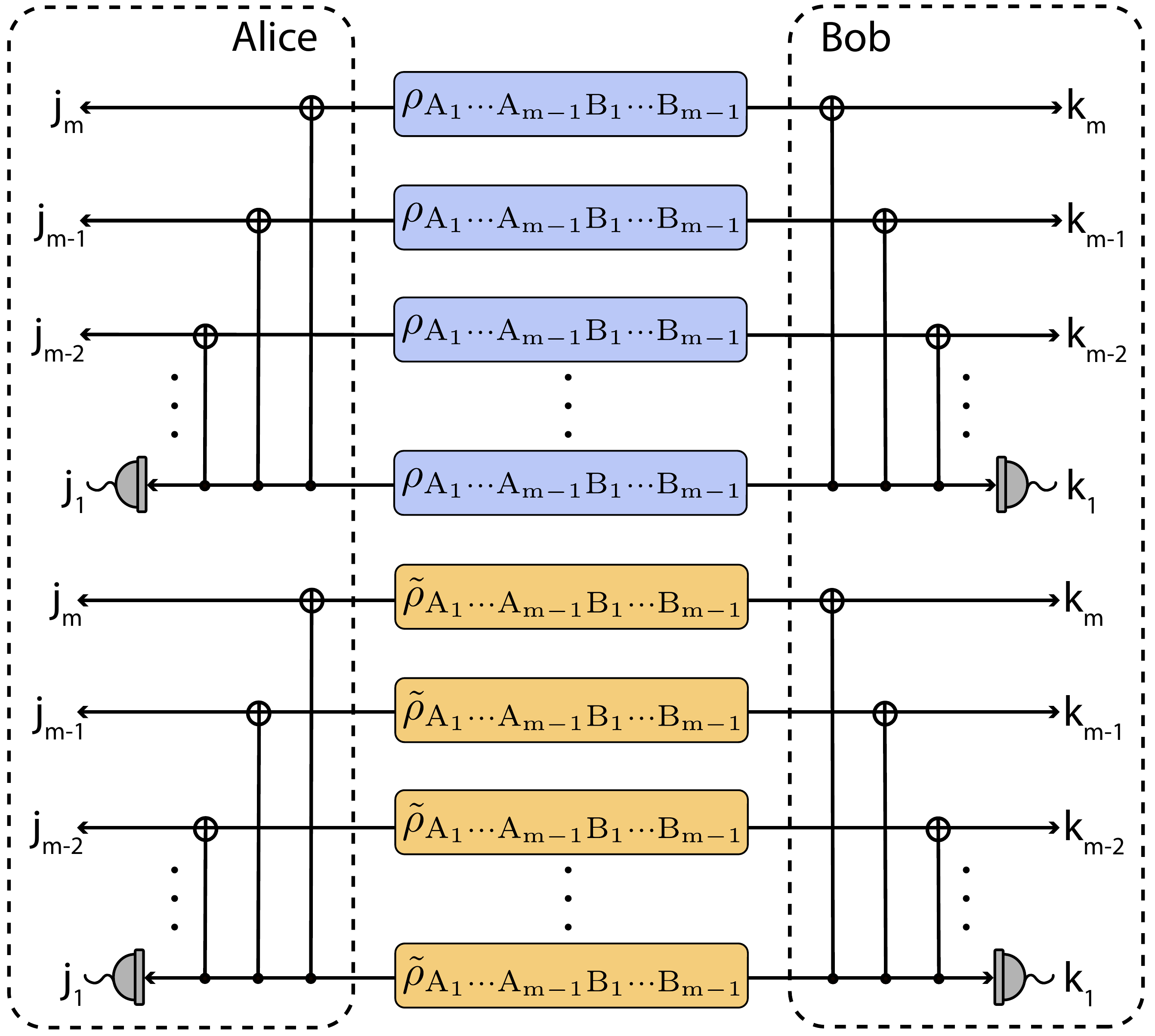}
\caption{\label{fig:stage2one}Second stage of the purification protocol, where blue and yellow boxes represent successful and failed states, respectively. The top half further purifies $m$ copies of the successfully purified states from the first round, while the bottom half purifies $m$ copies of the states that failed in the first round. The last rail serves as the control, with CNOTs applied and the last rail measured. If $j_1 = k_1$, further purification is achieved.}
\end{figure}

Figure~\ref{fig:stage1_result} presents the results of the first stage of the protocol. For $m=2$, the RCI, success probability, and fidelity of the output state in our protocol match the results of Deutsch's protocol~\cite{deutsch1996quantum}, despite the differing circuits. However, as the number of input Bell pairs increases, our protocol approaches the channel capacity, nearly saturating it at $m=30$. Our protocol extends Deutsch's recurrence protocol to the $N\rightarrow M$ regime providing an explicit circuit for the purification of any number of Bell pairs.
%Our protocol generalises Deutsch's protocol~\cite{deutsch1996quantum} to any number of Bell pairs, providing an explicit circuit for their purification. 
It's important to note that Deutsch's protocol~\cite{deutsch1996quantum} cannot saturate the channel capacity while purifying the Bell pairs, as each round results in half of the Bell pairs being discarded, corresponding to a ratio of $1/2^n$ of the capacity at each round $n$. 
This necessitates using more Bell pairs to reach the capacity while purifying. \ZJ{For instance, in our protocol, we discard only one Bell pair out of every $m$ pairs. Consequently, after each round of purification, we retain $(m-1)^n/m^n$ of the capacity, rather than $1/2^n$. However, as the number of Bell pairs increases, the rate of purification decreases, requiring additional rounds to achieve the desired purification level.}
%This necessitates using more Bell pairs to reach the capacity while purifying. However, as the number of Bell pairs increases, the rate of purification decreases, requiring more rounds of the purification process.

\subsection{\label{sec:second_stage}The Second Stage of the Protocol}
To further purify the Bell states at the dephasing channel capacity, we use $m$ copies of both the successfully purified high-dimensional states and the failed states from the first round, treating them separately. All the successful states are grouped together, and all the failed states are grouped with the other failed states as shown in figure~\ref{fig:stage2one}. CNOT gates are then applied between the last rail and each individual state, with the last rail serving as the control. The last rail is then measured and if $j_1 = k_1$, further purification is achieved.
%Equivalence between measuring one high-dimensional state and individual Bell pairs in the second round of the protocol, with $m=3$ from the first round. 
\begin{figure*}[t!]
%\hspace*{-0.3cm}
\includegraphics[scale=0.22]{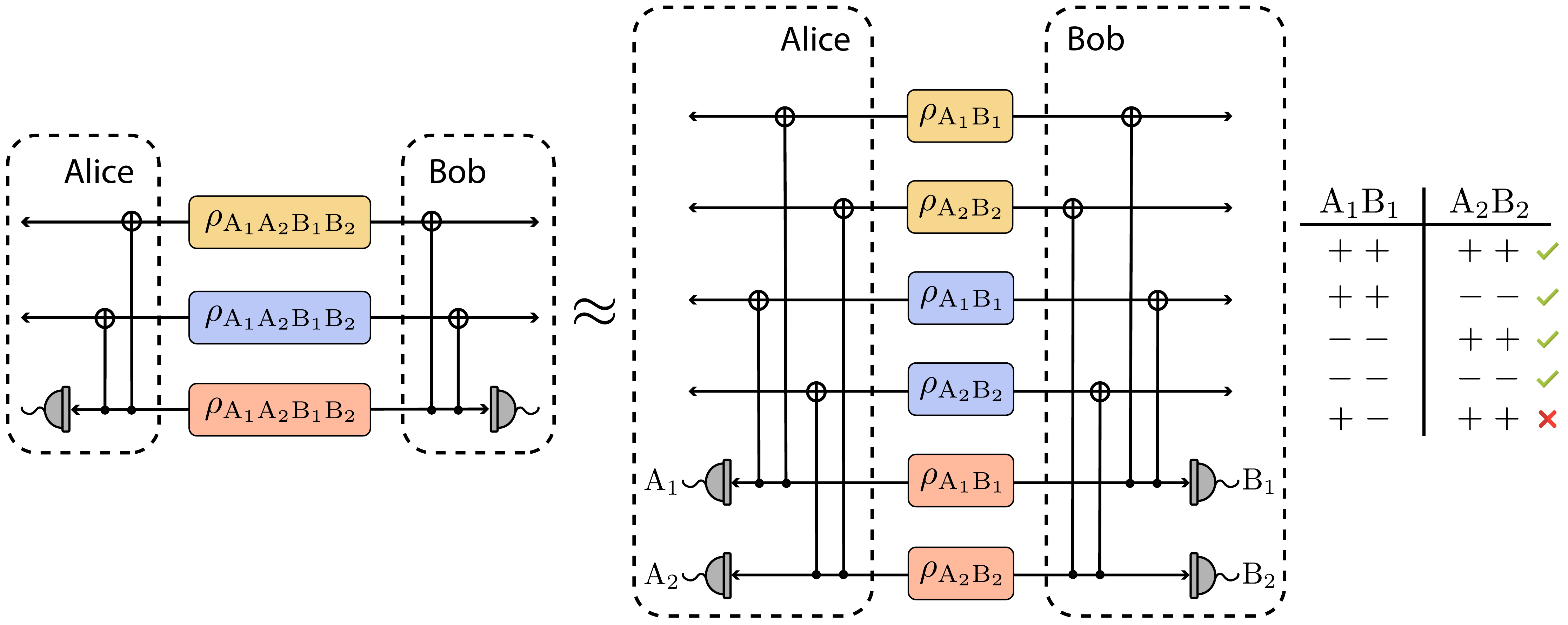}
\caption{\label{fig:stage2equal}\ZJ{Schematic representation of measuring a high-dimensional state in terms of individual Bell pairs in the second round of the protocol, with $m=3$ carried over from the first round.} In the second round, there are $m=3$ high-dimensional states $\rho_{\mathrm{A_1A_2B_1B_2}}$. CNOTs are applied between Bell pairs that are not entangled with each other. Each high-dimensional state contains two Bell pairs, so measuring one high-dimensional state corresponds to measuring the two Bell pairs within it, $\rho_{\mathrm{A_1B_1}}$ and $\rho_{\mathrm{A_2B_2}}$. Alice and Bob achieve further purification when $\mathrm{A_1}$ and $\mathrm{B_1}$ agree and $\mathrm{A_2}$ and $\mathrm{B_2}$ agree, corresponding to outcomes $(++,++), (++,--), (--,++)$ and $(--,--)$.}
\end{figure*}

Figure~\ref{fig:stage2equal} demonstrates the application of CNOTs to the high-dimensional states for $m=3$. The equivalent circuit is depicted in the center of figure~\ref{fig:stage2equal}. Following the first round of the protocol, the qubits within the high-dimensional state become entangled, with $\mathrm{A}_1$ entangled with $\mathrm{A}_2$ and $\mathrm{B}_1$ entangled with $\mathrm{B}_2$ in the state $\rho_{\mathrm{A_1A_2B_1B_2}}$. In the second round, CNOTs are applied between the control qubits and the qubits that are not entangled with each other. For instance, if $\rho_{\mathrm{A_1A_2B_1B_2}}$ is the control state, CNOTs are applied between qubit $\mathrm{A_1}$ of the control state and all the other $\mathrm{A_1}$ qubits, and similarly between qubit $\mathrm{A_2}$ of the control state and all the other $\mathrm{A_2}$ qubits. This avoids concentrating entanglement between the rails and instead focuses on enhancing the entanglement between Alice and Bob.
%within Alice’s or Bob’s own respective modes and instead focuses on enhancing the entanglement of the Bell pairs shared between them.

\begin{figure}[t!]
%\hspace*{-0.3cm}
\includegraphics[scale=0.3]{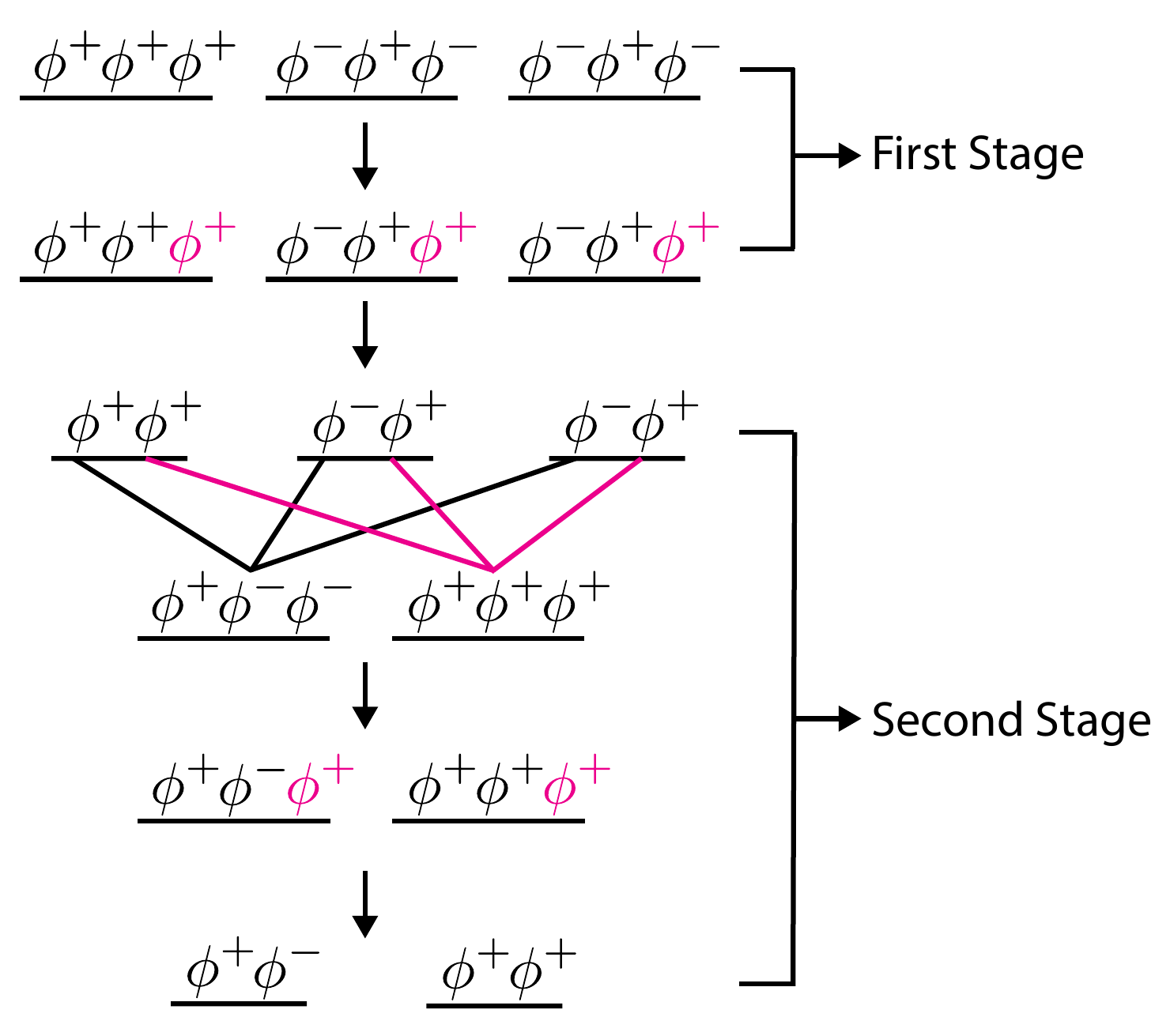}\hspace*{-0.2cm}
\caption{\label{fig:second_stage_shuffle_demo}
Illustration of the rearrangement of Bell pairs using a CNOT gate in the second stage of the protocol. Black and pink lines indicate how the Bell pairs are grouped together for the CNOT application in the second round.}
\end{figure}

The CNOTs applied in the second round function the same as in the first round. Suppose Alice distributes 9 Bell pairs to Bob, intending to perform purification over 2 rounds. Some Bell pairs may experience phase-flips from $\ket{\phi^+}$ to $\ket{\phi^-}$. For instance, with 4 transmission errors after the first stage, Alice and Bob share states $\phi^+\phi^+\phi^+$ and two $\phi^-\phi^+\phi^-$ with a probability of $(1-p)^5p^4$ in the first stage. Alice and Bob group these Bell pairs into three blocks of three and apply the circuit shown in figure~\ref{fig:stage2equal} (left). The first group $\phi^+\phi^+\phi^+$ has no errors, so the last Bell pair remains $\phi^+$ and is measured. The other two blocks $\phi^-\phi^+\phi^-$ each have 2 errors, converting the last Bell pair from $\phi^-$ to $\phi^+$ upon measurement, yielding a successful event. This is demonstrated in figure~\ref{fig:second_stage_shuffle_demo}. In the second stage, Alice and Bob 
apply the circuit shown in figure~\ref{fig:stage2equal} (center) with three blocks of two Bell pairs each: $\phi^+\phi^+$, $\phi^-\phi^+$, and $\phi^-\phi^+$. They regroup their qubits, combining all first pairs of each block into $\phi^+\phi^-\phi^-$ and $\phi^+\phi^+\phi^+$ as shown in figure~\ref{fig:second_stage_shuffle_demo}. The first block, with 2 errors, flips the last $\phi^-$ to $\phi^+$. The second block, with no errors, keeps $\phi^+$ as $\phi^+$. When both blocks have an even number of errors, the event is considered successful. 

As the state dimensions increase with each round of the protocol, simulating beyond $m=4$ in the second round becomes computationally demanding. Therefore, we simulate this round classically using truth tables and tracking errors in each block for Alice and Bob. The eigenvalues of the successful state of the second round, given success in the first round, are determined by
\begin{equation}
     \Tilde{\lambda}_{s_1s_2}(j)=(1-p)^{m^2-j}p^j,
     \label{eq:eigenvalues_second_success_success}
\end{equation}
where $j=0,4,6,\ldots,m(m-1)$ for odd $m$ and $j=0,4,6,\ldots,m^2-4,m^2$ for even $m$. For example, with $m=3$, the maximum number of errors per block to yield a successful event is 2, allowing up to 6 transmission errors out of 9 Bell pairs to still result in successful blocks. Similarly, for $m=4$, 16 Bell pairs are needed for two rounds of purification, with up to 16 errors permissible in the first stage across 4 blocks. However, note that for even $m$, the second highest number of transmission errors in the second stage is $j=m^2-4$ errors, as $j=m^2-2$ errors in the first stage can lead to odd errors in some blocks in the second stage. For instance, with $m=4$, one possible combination that $j=m^2-2=14$ transmission errors can occur in the first round is $\phi^-\phi^-\phi^-\phi^-$, $\phi^-\phi^-\phi^-\phi^-$, $\phi^-\phi^-\phi^-\phi^-$, and $\phi^-\phi^-\phi^+\phi^+$. Since each block has an even number of errors, this leads to success in the first round. However, when rearranged for the second round, new blocks become $\phi^-\phi^-\phi^-\phi^-$, $\phi^-\phi^-\phi^-\phi^-$, and $\phi^-\phi^-\phi^-\phi^+$. While the first two blocks have even errors, the last block has odd errors, leading to failure. 

Generalising the success probability of the second round as a function of $m$ and $p$ given the states were successful in the first round of the protocol is more challenging than equation~\eqref{eq:prob_success_stage1} as the the qubits are rearranged in the second round to concentrate the entanglement between Alice and Bob rather than the unwanted entanglement between the rails. However, the probability of success of the second round given that the states were also successful in the first round is given by
\begin{equation}
    P_{s_1s_2}(m)=\hspace{-1.5em}\sum_{j=4,6,\cdots,j_{m-1}}^{j_m}\hspace{-1.5em}M\,\frac{\Tilde{\lambda}_{s_1s_2}(j)}{P_{s_1}^m(m)},
    \label{eq:ps_of_second_round}
\end{equation}
where $j_{m}=m(m-1)$ and $j_{m-1}=m(m-1)-2$ if $m$ is odd or $j_m=m^2$ and $j_{m-1}=m^2-4$ if $m$ is even. $M$ represents the multiplicity of each eigenvalue $\Tilde{\lambda}_{s_1s_2}(j)$ and each eigenvalue needs to be normalised by $P_{s_1}^m(m)$ as the second stage of the protocol requires $m$ copies of the successful high-dimensional states from the first round of the protocol (see appendix~\ref{sec:appendix_probability_success} for a more generalised expression for the probability of success of the second round). Finally, the eigenvalues also need to be normalised by the probability of success of the second round to get the coefficients of various Bell state combinations in the final state as 
\begin{equation}
    \lambda_{s_1s_2}(j)=\frac{\Tilde{\lambda}_{s_1s_2}(j)}{P_{s_1}^m(m)P_{s_1s_2}(m)}.
    \label{eq:eigenvalue_second_success_success_normalised}
\end{equation}

Similar to the first stage, we need to keep conditional states that do not result in a successful event in this round, as they still contain useful entanglement. Discarding them would prevent us from saturating the channel capacity. Therefore, we need to compute the RCI of the failed states in the second stage, given that they succeeded in the first round. The eigenvalues of the failed state can be derived from the eigenvalues of the succeeding states in the first and second rounds. Specifically, for the dephasing channel, the average state in the second round, comprising both the successful and failed states given success in the first round, is equal to $m-1$ copies of the successful state from the first stage. This ensures that when combined, the overall state retains the same purity as $m-1$ copies of the original state from the first stage, matching its eigenvalues. This occurs because, after the first round, we obtain $m$ copies of a successful high-dimensional state, and measuring one of these copies does not alter the overall purity in the second round. However, among the conditional states, the successful state becomes purer while the failed state is less pure. The relationship between the average state from the second round of the protocol and $m-1$ copies of the successful state from the second round can be given as
\begin{equation}
    \rho^{\otimes (m-1)}_{\mathrm{AB_{s_1}}}=P_{s_1s_2}(m)\times\rho_{\mathrm{AB}_{s_1s_2}}+P_{s_1f_2}(m)\times\rho_{\mathrm{AB}_{s_1f_2}},
    \label{eq:link_between_first_and_second_stage}
\end{equation}
where $\rho_{\mathrm{AB}_{s_1}}$ represents the successful state from the first stage of the protocol. $P_{s_1s_2}(m)$ (given in equation~\eqref{eq:ps_of_second_round}) and $P_{s_1f_2}(m)$ are the probabilities of success and failure, respectively, in the second round given the state was successful in the first round. $P_{s_1f_2}(m)$ is equal to $1-P_{s_1s_2}(m)$. $\rho_{\mathrm{AB}_{s_1s_2}}$ and $\rho_{\mathrm{AB}_{s_1f_2}}$ are the successful and failed conditional states of the second round, respectively, given they succeeded in the first round. Using equation~\eqref{eq:link_between_first_and_second_stage}, the eigenvalues of the failed state can be expressed as
\begin{multline}
    \lambda_{s_1f_2}(j_1,j_2)=\frac{1}{P_{s_1f_2}(m)}\bigg[\frac{(1-p)^{m(m-1)-j_1}p^{j_1}}{P_{s_1}^{m-1}(m)}\\-\frac{(1-p)^{m^2-j_2}p^{j_2}}{P_{s_1}^m(m)}\bigg],
    \label{eq:eigenvalue_second_round_success_fail}
\end{multline}
where $j_2$ is the total number of transmission errors while $j_1$ is the number of remaining transmission errors after measuring some of the Bell pairs in the first round. The relationship between $j_1$ and $j_2$ is expressed as
\begin{subequations}
\begin{equation}
    \text{$m$=even}\begin{cases} 
     j_2=0 & j_1=0 \\
     j_2=4 & j_1=2 \\
     j_2=j_1, j_1+2, j_1+4 & 4\leq \!j_1<\!m(m-1)\!-\!2\\
     j_2=j_1, j_1+2 & j_1=m(m-1)-2\\
     j_2=m^2 & j_1=m(m-1),
   \end{cases} 
\end{equation}
\begin{equation}
    \text{$m$=odd}\begin{cases} 
     j_2=0 & j_1=0 \\
     j_2=4 & j_1=2 \\
     j_2=j_1, j_1+2, j_1+4 & 4\leq j_1<(m-1)^2\\
     j_2=j_1:m(m-1):2 & j_1=(m-1)^2.
   \end{cases} 
\end{equation}
\label{eq:boundary_eigenvalue_success}
\end{subequations}
\begin{figure}[t!]
%\hspace*{-0.3cm}
\includegraphics[scale=0.52]{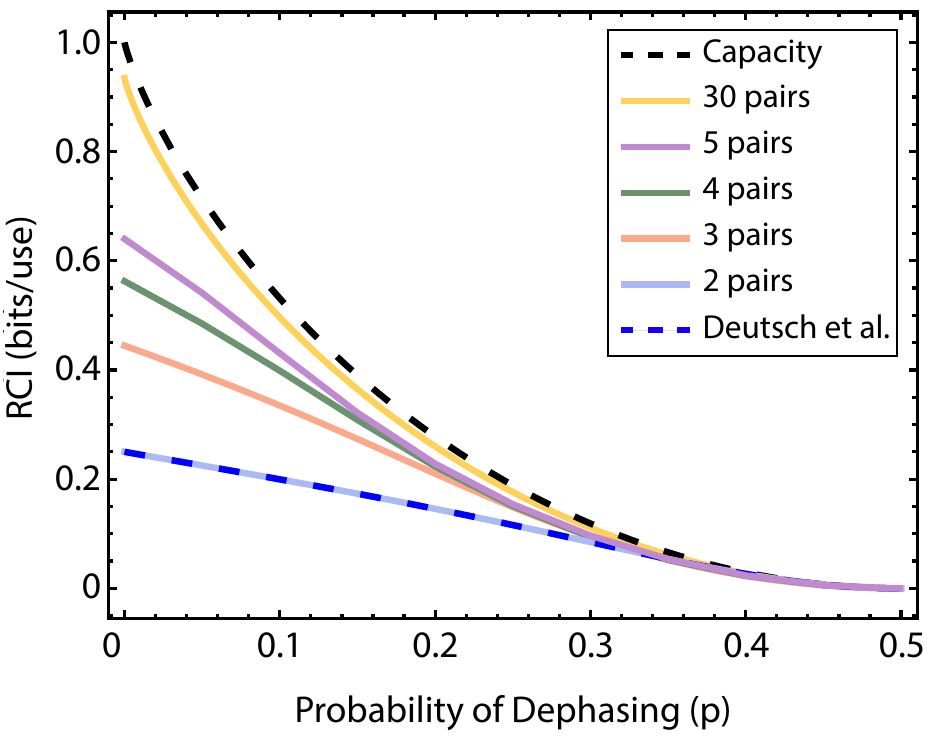}
\caption{\label{fig:stage2_result} Simulation results of the second round of the protocol with increasing number of Bell pairs $m$ shared between Alice and Bob. The black dashed line represents the dephasing channel capacity, while the blue dashed line shows results from Deutsch \textit{et al.}~\cite{deutsch1996quantum}. The yellow line for $m=30$ is simulated using $(m-1)^2/m^2\times C$ where $C$ represents the channel capacity.}
\end{figure}
Note that we do not discard the conditional failed states from the first round. Instead, we take $m$ copies of these high-dimensional failed states and attempt to purify them further in the second round. As mentioned earlier, to saturate the capacity, we need to retain all conditional states regardless of success or failure as they contain useful entanglement. Therefore, we compute the RCI of both successful and failed states in the second round, given they failed in the first round. To do this, we first compute their eigenvalues which are given in appendix~\ref{sec:appendix_eigenvalues_second_stage}. Then using all the conditional states, the RCI of the second round of the protocol can be written as
\begin{figure*}[hbt!]
%\hspace*{-0.3cm}
\includegraphics[scale=0.45]{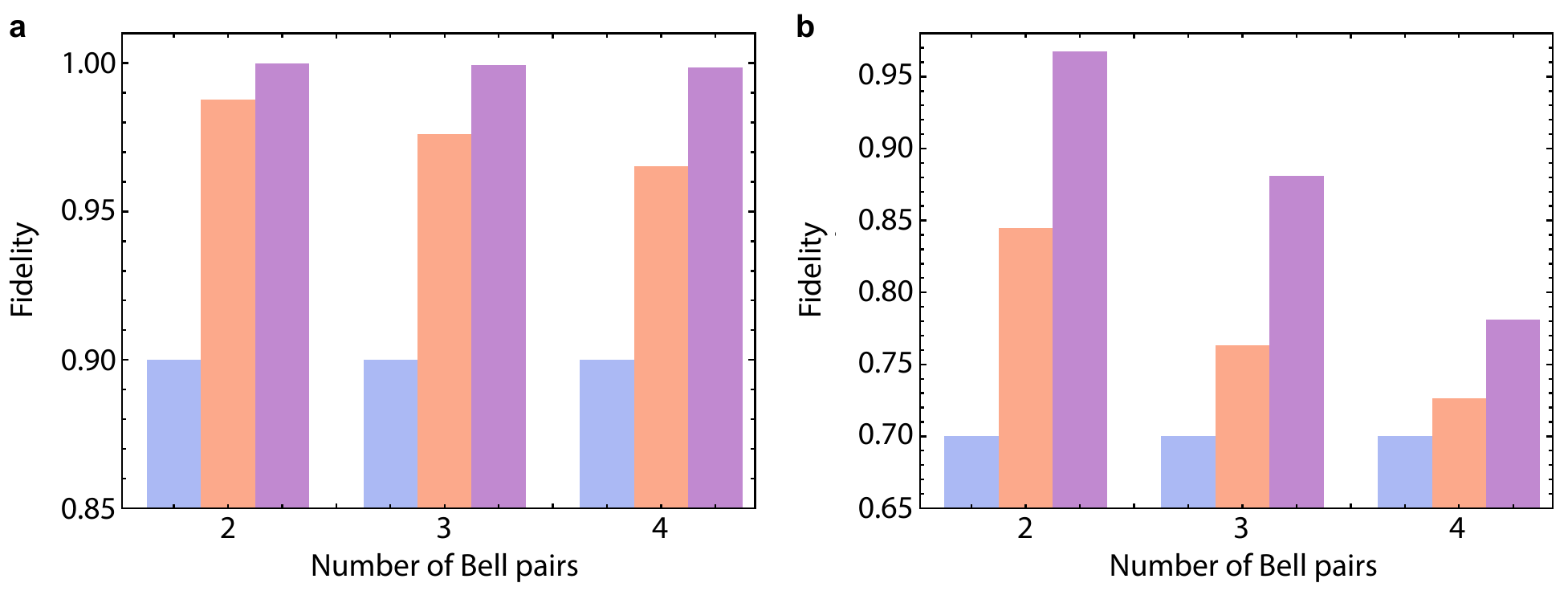}\hspace*{0.8cm}
\caption{\label{fig:fidelity_results} Comparison of the fidelity of the reduced successful states with the $\ket{\phi^+}$ state without purification and after the first and second rounds of the purification protocol. (a) presents the fidelity of the reduced states using $m=2$, $m=3$, and $m=4$ Bell pairs, with a dephasing probability of $p=0.1$, corresponding to an initial fidelity of $F=0.9$ before purification. Blue boxes indicate the fidelity prior to purification, while pink and purple boxes represent the fidelities after the first and second rounds, respectively. (b) shows the fidelity of the reduced states for an initial dephasing probability of $p=0.3$, corresponding to an initial fidelity of $F=0.7$.}
\end{figure*}
\begin{align}
    \label{eq:average_rci_stage2}
    \mathrm{RCI_2}=&\frac{1}{m^2}\big[P_{s_1}(m)\big(P_{s_1s_2}(m)\times\mathrm{RCI}(\rho_{\mathrm{AB}_{s_1s_2}}) \nonumber \\
    &+P_{s_1f_2}(m)\times\mathrm{RCI}(\rho_{\mathrm{AB}_{s_1f_2}})\big) \nonumber \\
    &+P_{f_1}(m)\big(P_{f_1s_2}(m)\times\mathrm{RCI}(\rho_{\mathrm{AB}_{f_1s_2}})\nonumber \\
    &+P_{f_1f_2}(m)\times\mathrm{RCI}(\rho_{\mathrm{AB}_{f_1f_2}})\big)\big],
\end{align}
where \ZJ{$\rho_{\text{AB}_{f_1s_2}}$ and $\rho_{\text{AB}_{f_1f_2}}$ represent the conditional states in the second round that succeeded and failed, respectively, given failure in the first round. The probabilities of success and failure in the second round, given first-round failure, are denoted by $P_{f_1s_2}(m)$ and $P_{f_1f_2}(m)$ (refer to appendix~\ref{sec:appendix_eigenvalues_second_stage} for how to obtain the eigenvalues and the probabilities).} Note that the average RCI is normalised by $m^2$ because distributing $m^2$ Bell pairs through the channel is required to proceed to the second round. The RCI of each conditional state is computed using equation~\eqref{eq:rci_equation}, where $S(\rho_\mathrm{A})$ for each conditional state is $\log_2(2^{(m-1)^2}) = (m-1)^2$, and $S(\rho_{\mathrm{AB}})$ is calculated by summing the eigenvalues, substituting each into $-\Tilde{\lambda}\log_2{\Tilde{\lambda}}$.

Figure~\ref{fig:stage2_result} illustrates the RCI during the second round of the protocol as $m$ increases. For a fixed value of $m$, each line shows \ZJ{an} expected decrease in performance compared to the first stage of the protocol, due to the normalisation of the average RCI by $1/m^2$. However, as $m$ increases, the average RCI converges towards the capacity of the dephasing channel. Notably, compared to the previous round, the fidelity of the successful reduced states improve \ZJ{as} shown in figure~\ref{fig:fidelity_results}. For $m=3$, the fidelity of the reduced two-mode state with $\ket{\phi^+}$ increases from $0.9$ to $0.9762$ in the first round of the protocol, reaching a fidelity of $0.9994$ after the second round, as shown in figure~\ref{fig:fidelity_results}(a). Similarly, for $m=4$, the fidelity starts at $0.9654$ in the first round and improves to $0.9986$ in the second round, demonstrating the protocol's capability to purify with an increasing number of rounds, as proven in the following section. It is important to note that as $m$ increases, the fidelity in each round will be slightly lower compared to that for a smaller value of $m$, since we only sacrifice one (high-dimensional) state each time each time. For example, measuring one state out of three contributes more to the remaining two than measuring one out of four. When the probability of dephasing increases from $p=0.1$ to $p=0.3$, corresponding to higher decoherence, our protocol still enhances the fidelities, as shown in figure~\ref{fig:fidelity_results}(b), albeit more gradually. This slower increase is expected due to the higher decoherence. However, when decoherence is high, fewer Bell pairs between Alice and Bob are needed to reach the channel capacity. Therefore, a smaller number of $m$ already saturates the channel capacity, as observed in both Figs.~\ref{fig:stage1_result} and~\ref{fig:stage2_result}.

%\section{\label{sec:optimality}Optimality of Our Protocol}
\subsection{\label{sec:achieving_capacity}Achieving the Capacity}
\begin{figure}[t!]
%\hspace*{-0.3cm}
\includegraphics[scale=0.45]{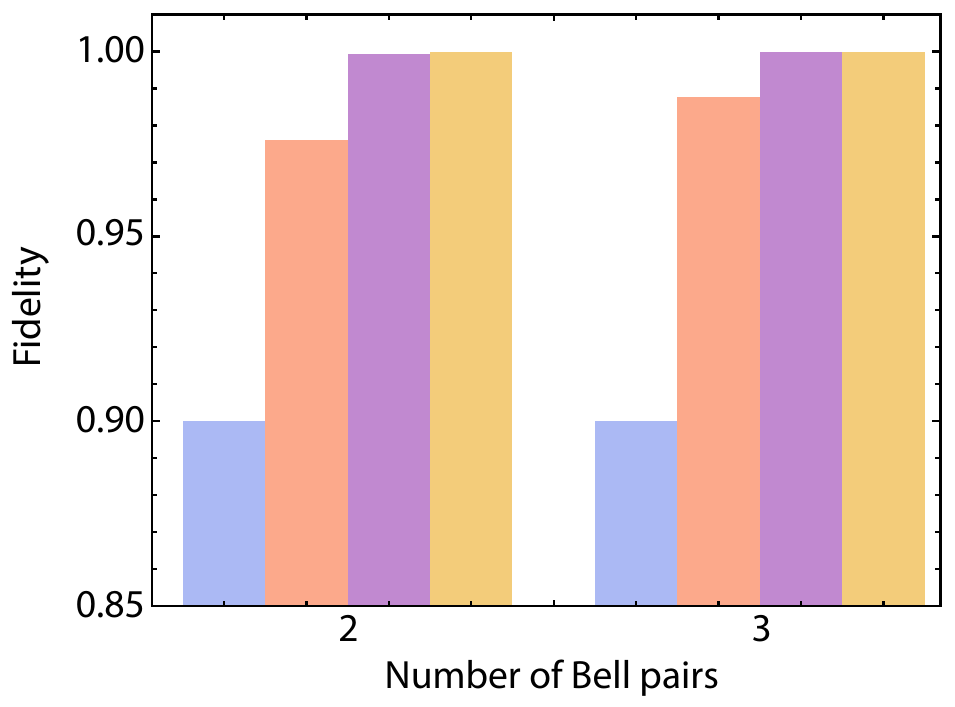}\hspace*{0.5cm}
\caption{\label{fig:fidelities_three_rounds} Comparison of the fidelity of the reduced states with the $\ket{\phi^+}$ state without purification and after the first, second and third rounds of the purification protocol. The left boxes are for $m=2$ while the right shows the results of $m=3$ for a dephasing value of $p=0.1$. Blue boxes indicate the fidelity prior to purification, while pink, purple and yellow boxes represent the fidelity after the first, second and third rounds, respectively.}
\end{figure}
As the dephasing channel is a distillable channel, we use RCI~\cite{pirandola2009direct, garcia2009reverse} shown in equation~\eqref{eq:rci_equation} as a benchmark to test the quality of our purification protocol. The RCI serves to quantify the amount of distillable entanglement between Alice and \ZJ{Bob} to infer the secret key rate. 
\begin{figure*}[hbt!]
%\hspace*{-0.3cm}
\includegraphics[scale=0.45]{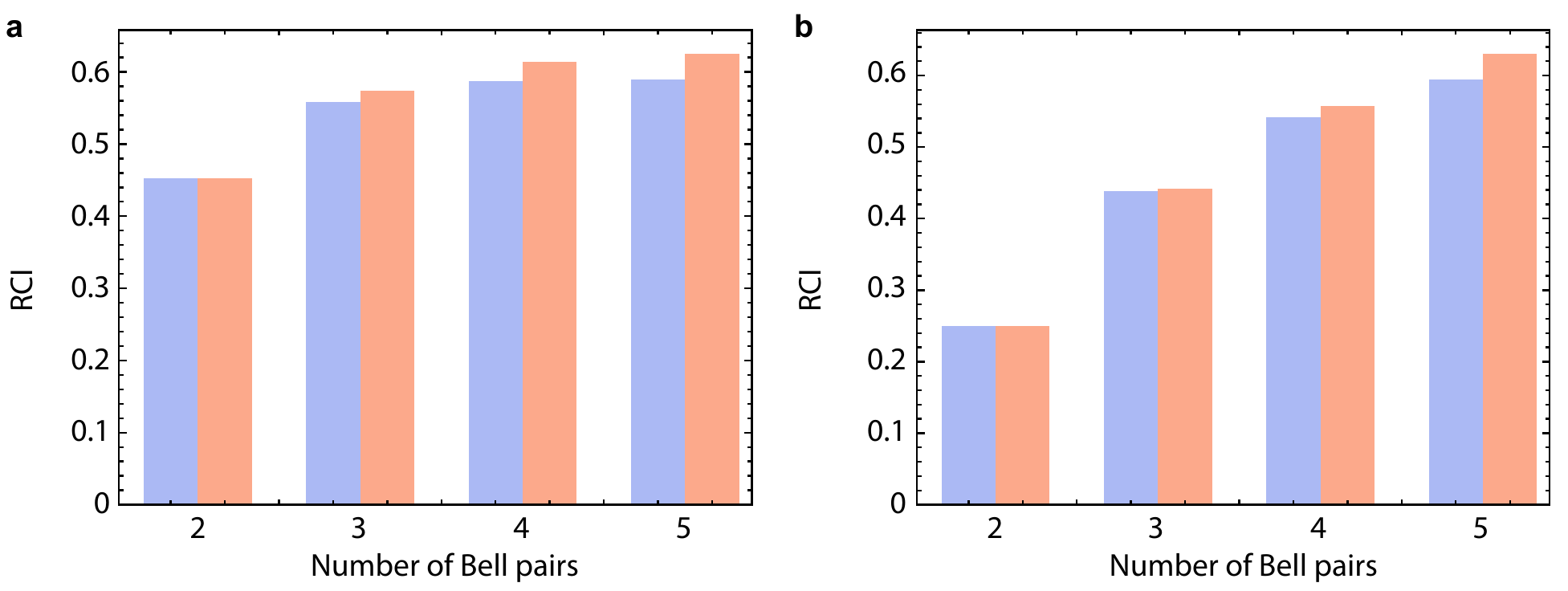}\hspace*{0.8cm}
\caption{\label{fig:rci_results_of_both_protocols} Comparison of the RCI between $m-1$ copies of the reduced successful states in the alternative protocol and the high-dimensional states in the actual protocol \ZJ{for a dephasing probability of $p=0.1$.} Blue boxes represent the alternative protocol, while pink boxes show the actual protocol. (a) illustrates the RCI of the states after the first round with an increasing number of Bell pairs, and (b) presents the results after the second round.}
\end{figure*}

In this section, we show no entanglement is lost between the rounds of the purification process as $m\rightarrow \infty$. If we add the successful and failed states up after the first round, they give the state of $m-1$ copies of the noisy Bell pairs as shown below
\begin{equation}
    \label{eq:rci_first_stage_relation}
    \frac{1}{m}\mathrm{RCI}(\rho^{\otimes (m-1)}_{\mathrm{AB}})\!\!=\!\!\frac{1}{m}\mathrm{RCI}(P_{s_1}\!(m)\times\rho_{\mathrm{AB}_{s_1}}\! \!+P_{f_1}\!(m)\times\rho_{\mathrm{AB}_{f_1}}\!),
\end{equation}
where $\rho_{\mathrm{AB}}^{\otimes(m-1)}$ is $m-1$ copies of the initial state shown in equation~\eqref{eq:choi}. The intuition behind the RHS of equation~\eqref{eq:rci_first_stage_relation} is that when combining the successful and failed states after the first round of the protocol, the overall state has the same entanglement as the state from the first stage with up to $m-1$ copies, since one copy was measured during the process. 
%the overall state should not be purified and should retain the same purity as the state from the first stage with up to $m-1$ copies, since one copy was measured during the process. 
Note that the RCI of $\rho_{\mathrm{AB}}$ is equal to $1-\mathrm{H}_2(p)$ which is the capacity, $C$ of the dephasing channel. As there is no entanglement between each copy of $\rho_{\mathrm{AB}}$ before the purification process, we can use $S(\rho^{\otimes n})=n\,S(\rho)$. Therefore the LHS of equation~\eqref{eq:rci_first_stage_relation} becomes
\begin{align}
    \label{eq:rci_first_stage_relation_LHS}
    \frac{1}{m}\mathrm{RCI}(\rho^{\otimes (m-1)}_{\mathrm{AB}})=&\frac{1}{m}\,\big(S(\rho_{\mathrm{A}}^{\otimes(m-1)})-S(\rho_{\mathrm{AB}}^{\otimes(m-1)})\big)\nonumber \\
    &=\frac{m-1}{m}\big(S(\rho_{\mathrm{A}})-S(\rho_{\mathrm{AB}})\big) \nonumber \\
    &=\frac{m-1}{m}(1-\mathrm{H_2}(p)) \nonumber \\
    &=\frac{m-1}{m}C.
\end{align}
This implies that after the first round of the protocol, the average state has the same RCI as $m-1$ copies from the previous round (see appendix~\ref{sec:appendix_cap_first} and appendix~\ref{sec:appendix_cap_second} for the proof of the RHS of equation~\eqref{eq:rci_first_stage_relation} and the capacity proof for the second stage, respectively). Additionally, as $m\rightarrow \infty$,
\begin{equation}
    \label{eq:infinity_RCI_relation_first round}
    \lim_{m\to\infty}\frac{m-1}{m}C=C.
\end{equation}

For the $n^\text{th}$ round of the protocol, the RCI of the average state between Alice and Bob can be expressed as 
\begin{equation}
    \label{eq:RCI_rounds}
    \mathrm{RCI}(\rho_{\mathrm{AB}_{n}})=\frac{1}{m^n}\mathrm{RCI}(\rho_{\mathrm{AB}_{n-1}}),
\end{equation}
where $\mathrm{RCI}(\rho_{\mathrm{AB}_{n}})$ denotes the RCI of the average state of the current round, while $\mathrm{RCI}(\rho_{\mathrm{AB}_{n-1}})$ represents the RCI of $m-1$ copies of the average state from the previous round. The overall RCI needs to be normalised by the total number of Bell pairs used throughout the purification proceess. Since the RCI of \ZJ{the current} round is equal to the previous round, the RCI of the $n^{\text{th}}$ can be related to the first round by
\begin{equation}
    \label{eq:RCI_rounds_relation_to_round_one}
    \mathrm{RCI}(\rho_{\mathrm{AB}_{n}})=\frac{1}{m^n}\mathrm{RCI}(\rho_{\mathrm{AB}}^{\otimes (m-1)^n}).
\end{equation}
Using the same relations as equation~\eqref{eq:rci_first_stage_relation_LHS}, equation~\eqref{eq:RCI_rounds_relation_to_round_one} becomes
\begin{align}
    \label{eq:RCI_capacity_any_round}
    \mathrm{RCI}(\rho_{\mathrm{AB}_{n}})&=\bigg(\frac{m-1}{m}\bigg)^n\mathrm{RCI}(\rho_{\mathrm{AB}})\nonumber \\
    &=\bigg(\frac{m-1}{m}\bigg)^n\,C.
\end{align}
If $n=m$, the limit as $m\rightarrow\infty$ approaches \ZJ{$C/e$}, which would result in the lower bound being less than the capacity. However, in each round, the effective dephasing is reduced compared to the previous round, leading to higher distillable entanglement. This might suggest a violation of the capacity. Nonetheless, this is prevented by the coefficient $\big(\frac{m-1}{m}\big)^n$ that appears before the capacity. Therefore, as long as $n\ll m$, we have 
\begin{equation}
    \label{eq:proof_of_capacity_for_any_n}
    \lim_{m\to\infty}\bigg(\frac{m-1}{m}\bigg)^n\,C=C. 
\end{equation}
\begin{figure}[h!]
\hspace*{-0.3cm}
\includegraphics[scale=0.52]{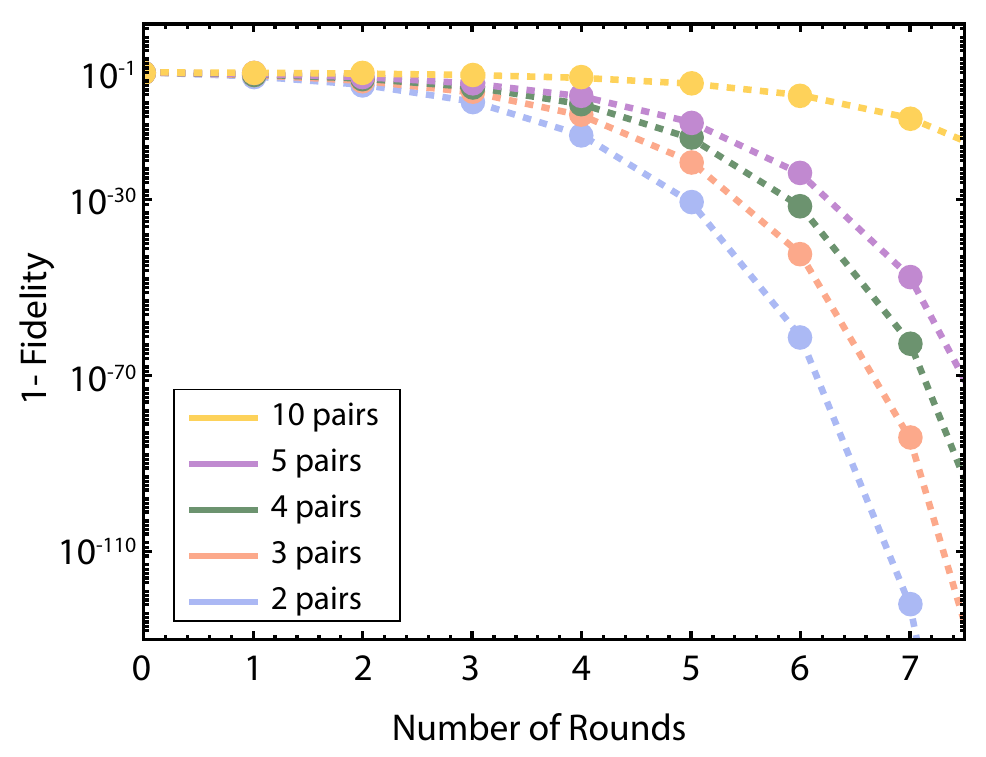}\hspace*{0.1cm}
\caption{\label{fig:fidelities_approximate_results}
Dephasing probability of the alternative protocol (also corresponding to 1-fidelity) as a function of rounds with increasing number of Bell pairs. The initial dephasing probability is $p=0.1$ corresponding to an initial fidelity $F=0.9$ with respect to $\ket{\phi^+}$ state.}
\end{figure}

For $p=0.1$ and $m=2$ and $m=3$, a fidelity very close to $1$ is achieved within 3 rounds for both cases, as illustrated in figure~\ref{fig:fidelities_three_rounds}. Although it is not immediately obvious in the figure, both fidelities approach $1$ by the third round, starting from $F_2=0.9998$ for $m=2$ and $F_2=0.9994$ for $m=3$. This suggests that the rate at which the number of rounds needs to increase as $m$ grows may not scale as quickly as $m$ itself to purify while achieving the capacity.

\ZJ{For more rigorous proofs on achieving the channel capacity, we show in appendix~\ref{sec:appendix_cap_first} that the right-hand side of equation~\eqref{eq:rci_first_stage_relation} equals the left-hand side in the first round of the protocol, with an alternative proof provided by lower bounding the RCI in equation~\eqref{eq:average_rci_stage1}. This is further used in appendix~\ref{sec:appendix_cap_second}, where we lower bound the RCI in the second round using equation~\eqref{eq:average_rci_stage2} to confirm that it achieves the capacity. Finally, in appendix~\ref{sec:appendix_any_round}, we extend the proof to show that the protocol can achieve the channel capacity in any round of the purification process.}

\subsection{Purifying the States}
%While our protocol purifies the dephased Bell states, it also generates entanglement between Alice's and Bob's respective modes. 
%\textcolor{red}{While our protocol purifies the dephased Bell states, it also generates nontrivial global correlations across the purified qubits.} For instance, for $m=3$, after the first round we obtain the high-dimensional state $\rho_{\mathrm{A_1B_1A_2B_2}}$ with reduced states $\rho_{\mathrm{A_1B_1}}$ and $\rho_{\mathrm{A_2B_2}}$ which are both individually purer and identical. \textcolor{red}{Although the marginal states such as $\rho_{\mathrm{A_1A_2}}$ and $\rho_{\mathrm{B_1B_2}}$ are maximally mixed, the overall state is not a tensor product of independent Bell pairs (e.g. $\rho_{\mathrm{A_1B_1A_2B_2}}\neq \rho_{A_1B_1}^{\otimes2}$). Therefore, these global correlations cannot be disregarded, as they contribute to attaining the channel capacity and further purification of the Bell pairs in subsequent rounds.}
While our protocol purifies the dephased Bell states, it also generates entanglement between Alice’s and Bob’s respective modes. For instance, for $m=3$, after the first round we obtain the high-dimensional state $\rho_{\mathrm{A_1B_1A_2B_2}}$ with reduced states $\rho_{\mathrm{A_1B_1}}$ and $\rho_{\mathrm{A_2B_2}}$ which are both individually purer and identical. However, $\mathrm{A_1}$ and $\mathrm{A_2}$ now exhibit correlations that cannot be disregarded, as this entanglement contributes to attaining the channel capacity and further purification of the Bell pairs in subsequent rounds.

To prove that our protocol achieves the required fidelity of $F=1$ and fully purifies the states, we can derive a lower bound. To this end, we propose an alternative protocol in which the correlations between Alice's and Bob's own modes are disregarded. In this alternative approach, instead of using $m$ copies of $\rho_{\mathrm{A_1B_1A_2B_2}}$, we consider $m$ copies of the reduced state $\rho_{\mathrm{A_1B_1}}$ and assume that the protocol is restarted from the first round with these fresh copies using the circuit in figure~\ref{fig:stage1}, which are much purer than the initial $\rho_{\mathrm{AB}}$. Proving that the alternative protocol serves as a lower bound to the actual protocol is challenging because finding a general expression for any number of copies, $m$, and rounds, $n$, is intractable due to the complexity of the eigenvalue multiplicities. However, figure~\ref{fig:rci_results_of_both_protocols} provides numerical evidence that the RCI of the successful states in the alternative protocol is consistently lower than that of the actual protocol. Recall that RCI quantifies the distillable entanglement. As shown in figure~\ref{fig:rci_results_of_both_protocols}(a), the RCI of $m-1$ copies of the reduced state is lower than that of the high-dimensional state obtained after the first round of the protocol, indicating that some useful entanglement between the rails is discarded. This effect is similarly observed in figure~\ref{fig:rci_results_of_both_protocols}(b) after the second round. %Therefore, the entanglement disregarded by the alternative protocol is valuable and contributes to further concentrating the entanglement between Alice and Bob in subsequent rounds.

\ozlem{The relationship between the fidelities of the two protocols can be further understood through the quantum data processing inequality (DPI), a key principle in quantum information theory. The DPI states that applying a completely positive trace-preserving (CPTP) map such as a local operation, lossy transformation, or quantum channel, to a quantum state cannot increase its fidelity with respect to a reference state~\cite{lindblad1975completely, cover1999elements, beaudry2011intuitive, beigi2013new}.
%processing a quantum state through any local operation, lossy transformation, or quantum channel cannot increase its fidelity with respect to a reference state~\cite{lindblad1975completely, cover1999elements, beaudry2011intuitive, beigi2013new}. 
%In this context, the alternative protocol can be viewed as introducing a lossy process, where useful correlations between modes are discarded. Therefore,
In the alternative protocol, the process of tracing out certain modes can be regarded as a lossy operation that discards useful correlations between subsystems. Therefore, this can be expressed as
\begin{equation}
    F^A(\rho_{\text{AB}_{s_1}},\ket{\phi^+})\leq F^R(\rho_{\text{AB}_{s_1}},\ket{\phi^+}),
\end{equation}
where $F^A$ and $F^R$ denote the fidelities of the alternative and the original protocols with respect to the $\ket{\phi^+}$ state, respectively. 
%Here, the loss of entanglement acts as an additional "processing" step that degrades the state. This explains why the fidelity of the alternative protocol is slightly lower than that of the original protocol. Intuitively, the discarded entanglement in the alternative protocol is analogous to removing details from an image during compression. While the result remains close to the original, it lacks the fidelity achieved when all correlations are discarded.
}

\ozlem{It is worth noting that DPI is often discussed in contexts where the fidelity between two quantum states increases after processing because the operation "blurs" their differences. However, in this work, fidelity is measured with respect to a fixed reference state ($\ket{\phi^+}$). The partial trace operation, as a CPTP map, degrades the state relative to this fixed reference, rather than increasing fidelity as seen in other DPI contexts.
}

\ozlem{This fidelity degradation can be intuitively understood by considering the discarded subsystems as an additional "processing step" that weakens the entanglement in the remaining state. An analogy to image compression illustrates this: removing certain details during compression results in an image that remains recognisable but lacks the full fidelity of the original.}

%After the first round of the protocol, the reduced states have the following form
\ozlem{To quantify this effect, we examine the reduced states after the first round of the protocol}
\begin{equation}
    \Tilde{\rho}_{\mathrm{AB}}=\begin{pmatrix}
    0.5 & 0 & 0 & b/2a \\
    0 & 0 & 0 & 0 \\
    0 & 0 & 0 & 0 \\
    b/2a & 0 & 0 & 0.5
    \end{pmatrix},
    \label{eq:reduced_density_matrix}
\end{equation}
where before the purification protocol, the off-diagonal term is $b/2a=(1-2p)/2$. However, after the purification process, the new value of $a$ and $b$ becomes
\begin{subequations}
    \begin{equation}
        a(p)=2^{m-2}\bigg[\bigg(\frac{1-2p}{2}\bigg)^m\!+\bigg(\frac{1}{2}\bigg)^m\bigg],
        \label{eq:value_of_a}
    \end{equation}
    \begin{equation}
        b(p)=2^{m-2}\bigg[\!\bigg(\frac{1-2p}{2}\bigg)^{m-1}\!\bigg(\frac{1}{2}\bigg)+\bigg(\frac{1-2p}{2}\bigg)\!\bigg(\frac{1}{2}\bigg)^{m-1}\bigg].
        \label{eq:value_of_b}
    \end{equation}
    \label{eq:values_of_a_and_b}
\end{subequations}
Using equations~\eqref{eq:value_of_a} and~\eqref{eq:value_of_b}, the new dephasing probability becomes
\begin{equation}
\label{eq:new_dephasing_probability}
    \Tilde{p}(p)=\frac{1}{2}\bigg(1-\frac{b(p)}{a(p)}\bigg),
\end{equation}
and the fidelity of the new states is given by $F=1-\Tilde{p}$. It is important to note that the fidelities of the alternative protocol are always lower than those of the actual protocol but are still an improvement over the original states as depicted in table~\ref{tab:table_fidelity_comparison}. %This is because the alternative protocol ignores the beneficial entanglement between the rails, which is harnessed in the actual protocol to achieve higher fidelities.
\begin{table}[t]
\caption{\label{tab:table_fidelity_comparison}
Fidelities of the alternative protocol and the original protocol over $n=3$ rounds for $m=3$.
}
\begin{ruledtabular}
\begin{tabular}{ccc}
Round&
Alternative Protocol&
Original Protocol\\
\colrule
 $F_0$ & 0.9 & 0.9 \\
 $F_1$ & 0.97619 & 0.97619 \\
 $F_2$ & 0.998812 & 0.999384 \\
 $F_3$ & 0.999997 & 1 \\
\end{tabular}
\end{ruledtabular}
\end{table}

As this alternative protocol has slightly worse fidelity due to ignoring the useful entanglement between the respective modes, the relationship between the fidelities can be expressed as
\begin{subequations}
\label{eq:purity_relations_with_approximated_protocols}
\begin{equation}
    1-p\leq F^A\leq F^R\leq 1,    
    \label{eq:fidelity_relationship}   
\end{equation}
\begin{equation}
    0\leq p^R \leq p^A \leq p \leq \frac{1}{2},
    \label{eq:probability_relationship}
\end{equation}
\end{subequations}
where $p^A$ and $p^R$ denote the probabilities, of the alternative and the \ZJ{original} protocols, respectively.

To demonstrate that our protocol fully purifies the states, we use an iterative map proof, establishing that $\Tilde{p}^A\leq p\leq 1/2$ after the first round of the alternative protocol for any $m$. Following the second round, the relationship $\Tilde{p}^{A'}\leq \Tilde{p}^A\leq p$ holds, where $\Tilde{p}^{A'}$ and $\Tilde{p}^A$ represent the new probabilities after the second and first rounds of the alternative protocol, respectively. This indicates that with each repetition of the protocol, the probability of dephasing continually decreases, resulting in further purification of the states. The details of this proof are provided in appendix~\ref{sec:appendix_fidelity}. Additionally, we present numerical evidence in figure~\ref{fig:fidelities_approximate_results} showing that the approximate protocol closely mirrors our original protocol, achieving complete purification of the states over multiple rounds for any number of Bell pairs, $m$. The key difference between the two protocols is that the approximate protocol, by ignoring some useful entanglement, purifies the states slower than the original protocol.

\subsection{Cost of Purification}
\ozlemREV{While our protocol achieves the dephasing-channel capacity only in the asymptotic limit, it is important to note that the same circuit can be applied to a finite number of input states. In practice, if the aim is to generate a desired number of high-fidelity Bell pairs rather than to saturate the capacity, the protocol can be repeated for a finite number of rounds. In this case, the residual dephasing probability after $n$ rounds for small $p$ is given by (refer to Appendix~\ref{sec:appendix_cost_of_purification} for the derivation) 
\begin{equation}
    \label{eq:new_dephasing_prob_round_n}
    \Tilde{p}_n\approx p^{2^n}(m-1)^{2^n-1}+\mathcal{O}(p^{2^n+1}).
\end{equation}
Neglecting higher-order terms $\mathcal{O}(p^{2^n+1})$, the number of rounds required to achieve a fidelity $F = 1 - \delta$ is
\begin{align}
    \label{eq:round_vs_delta_fidelity}
    n&=\log_2\bigg(\frac{\log\big(\delta(m-1)\big)}{\log\big(p(m-1)\big)}\bigg)\nonumber \\
    &\approx\mathcal{O}\big(\log(|\log\delta|)\big)
\end{align}
where $\delta$ is the remaining dephasing error in the states, and the corresponding yield for this target $\delta$ at any value of $n$ is given by
\begin{equation}
    \label{eq:yield}
    Y_n=\bigg(\frac{m-1}{m}\bigg)^n.
\end{equation}
By substituting Eq.~\eqref{eq:round_vs_delta_fidelity} into Eq.~\eqref{eq:yield}, the yield can be lower bounded by
\begin{align}
    \label{eq:yield_exponential}
    Y_n &\geq 1 - \frac{n}{m}\nonumber \\
    Y_n &\geq 1 - \mathcal{O}\bigg(\frac{\log(|\log \delta|)}{m\log 2}\bigg),
\end{align}
%\begin{align}
%    \label{eq:yield_exponential}
%    Y_n&=\exp\bigg[n\log\bigg(\frac{m-1}{m}\bigg)\bigg]\nonumber \\
%    &=\exp\bigg[\log\bigg(\frac{m-1}{m}\bigg)\log_2\bigg(\frac{\log\big(\delta(m-1)\big)}{\log\big(p(m-1)\big)}\bigg)\bigg]
%    %&\approx \exp\big[\mathcal{O}\big(\log(|\log\delta|)\big)\big]
%\end{align}
%Thus, as $\delta \to 0$, the yield depends only on $\log(\log|\delta|)$, indicating that high fidelities can be reached with only a very mild reduction in yield, which is displayed in Table~\ref{tab:table_cost_of_purification}.
Thus, for fixed $\delta$ and $p$, Eq.~\eqref{eq:yield_exponential} shows that the gap $1 - Y_n$ is at most of order  $\mathcal{O}\big((\log(|\log \delta|)/(m\log 2)\big)$. Consequently, increasing $m$ drives the yield closer to one, with only a very weak (double-logarithmic) dependence on $\delta$, as illustrated in Table~\ref{tab:table_cost_of_purification}.
}
\begin{table}[t]
\caption{\label{tab:table_cost_of_purification}
Cost of purification for various target errors $\delta$ at an initial dephasing value of $p=0.1$. 
For each $\delta$, we list the required number of rounds $n$ and corresponding yield $Y$ for $m=3,4,5,8$.}
\begin{ruledtabular}
\begin{tabular}{c|cc|cc|cc|cc}
\multirow{2}{*}{$\delta$} 
& \multicolumn{2}{c|}{\makecell[c]{$m = 3$}}
& \multicolumn{2}{c|}{\makecell[c]{$m = 4$}}
& \multicolumn{2}{c|}{\makecell[c]{$m = 5$}}
& \multicolumn{2}{c}{\makecell[c]{$m = 8$}}
\\
& $n$ & $Y$ & $n$ & $Y$ & $n$ & $Y$ & $n$ & $Y$ \\
\colrule
$10^{-2}$--$10^{-3}$     & 2 & 0.44 & 2 & 0.56 & 2 & 0.64 & 3 & 0.67 \\
$10^{-5}$--$10^{-7}$     & 3 & 0.30 & 4 & 0.32 & 4 & 0.41 & 5 & 0.51 \\
$10^{-10}$--$10^{-12}$   & 4 & 0.20 & 5 & 0.24 & 5 & 0.33 & 6 & 0.45 \\
\end{tabular}
\end{ruledtabular}
\end{table}

\ozlemREV{Therefore, these results highlight two key aspects of the protocol. Equation~\eqref{eq:new_dephasing_prob_round_n} shows that each round induces a quadratic reduction in the dephasing noise, while the overall fidelity increases doubly-exponentially, with the error scaling as $p^{2^n}$. As quantified in Eq.~\eqref{eq:round_vs_delta_fidelity}, this double-exponential decay implies that very high fidelities can be achieved with only a small number of rounds $n\approx\mathcal{O}\big(\log(|\log\delta|)\big)$. Since this analysis is based on the simplified recursion of Eq.~\eqref{eq:new_dephasing_probability}, the full protocol exhibits even faster convergence.}

\section{\label{sec:conclusions}Conclusions}
Quantum communications enable secure information transfer and entanglement distribution but are limited by channel capacity and decoherence. Developing protocols that mitigate these challenges is key to realising quantum networks.

The Pauli dephasing channel exemplifies how decoherence can be particularly detrimental. While quantum repeaters can help mitigate losses and improve secret key rates beyond those of point-to-point communications, they fail to surpass the capacity of a repeaterless channel in the case of dephasing noise. This is because the Bell state measurements in repeaters further degrade the quantum states. To overcome this problem, we presented a purification protocol assisted with two-way classical communications designed to reach the dephasing channel capacity in the asymptotic limit. Through a recursive process, where Alice and Bob apply CNOT gates and measure in the Hadamard basis, the protocol purifies Bell pairs, improving their fidelity by correcting higher-order dephasing errors with each round. Ultimately, this method ensures the Bell pairs are nearly perfect, making them ideal for use in quantum repeaters before entanglement swapping with the respective nodes. Additionally, our protocol can also be applied to dephasing errors in quantum computers or to purify Bell states in distributed quantum computing. While achieving the dephasing-channel capacity requires multiple Bell pairs, our protocol offers an explicit circuit that can be adapted to any number of pairs, with the repetition rate adjustable to the user’s requirements. Importantly, our protocol remains relevant even outside the asymptotic limit. When implemented with small values of $m$, it naturally reduces to well-known schemes such as Deutsch’s protocol~\cite{deutsch1996quantum} and can be practically adapted into pumping~\cite{dur1999quantum} or banding-style~\cite{riera2021entanglement} purification strategies. Pumping protocols repeatedly distill a slightly purified state using fresh noisy states, while banding protocols group intermediate states by fidelity and apply tailored purification steps within each group. Due to our protocol's optimisation for the dephasing channel, the same fixed circuit can be reused across rounds, providing a simple and scalable approach that performs well even under realistic resource constraints.

\ozlemREV{Beyond these physical-level purification strategies, recent work has introduced constant-rate and constant-overhead distillation protocols based on stabiliser and quantum error-detecting codes, primarily in the context of fault-tolerant distributed quantum computing~\cite{pattison2025constant,bonilla2025constant,shi2025stabilizer}. These schemes operate on large encoded blocks and rely on syndrome measurements to reduce overhead. In contrast, our protocol acts directly on unencoded physical Bell pairs and is analytically optimised for the Pauli dephasing channel, providing an explicit circuit that approaches the channel’s two-way capacity. The two approaches therefore target different operational regimes: code-based schemes optimise fault-tolerant, encoded entanglement generation, whereas our protocol provides a physical-level, channel-specific purification routine.}

Although our circuit is specifically designed for the dephasing channel, it provides a framework for addressing noise and mitigating decoherence. As such, it can serve as a guide for adapting circuits to other noisy channels, such as depolarising, erasure, and thermal noise channels, where customised designs are required. Future work should also evaluate the protoocol's performance under more realistic conditions, including gate errors, memory decoherence, and compound noise, and explore extensions through circuit refinements.

\section*{Data Availability Statement}
The data that support the findings of this study are available upon reasonable request from the authors.

\section*{Acknowledgements}
I would like to thank my partner, Zachary Wood, for his invaluable support and insights in identifying patterns within the protocol, which contributed to this work. We also extend our gratitude to the Australian Government through the Australian Research Council Centre of Excellence for Quantum Computation and Communication Technology (Grant No. CE170100012) and to A$^*$STAR for their support through grants C230917010 (Emerging Technology), C230917004 (Quantum Sensing), and Q.InC Strategic Research and Translational Thrust.
\begin{figure}[t]
%\hspace*{-0.3cm}
\includegraphics[scale=0.21]{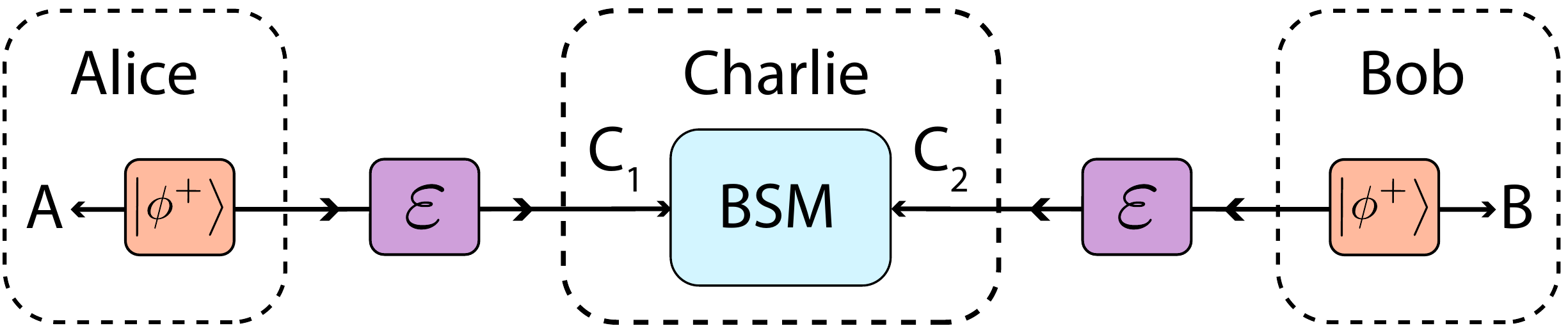}\hspace*{-0.01cm}
\caption{\label{fig:bsm_charlie} 
Alice and Bob each prepare a $\ket{\phi^+}$ state, keeping one qubit and sending the other to Charlie. Charlie then interferes with the two incoming qubits and performs a Bell-state measurement, establishing entanglement between Alice and Bob.
}
\end{figure}

%\section{Appendix}
\appendix
\section{\label{sec:appendix_repeater_proof}Proof of the Ineffectiveness of Bell-State Measurements in Pauli Dephasing Channel}
In this section, we demonstrate that, even with the inclusion of a node between the trusted parties, an entanglement swapping process using Bell-state measurements results in a state identical to that of a repeaterless channel. The state in equation~\eqref{eq:choi_repeater} represents the states shared between Alice and Charlie, and Charlie and Bob, after both Alice and Bob send a state to Charlie. When the original dephasing channel is divided into two by introducing a node, the Kraus operators for the channel with a node are given as
\begin{subequations}
\label{eq:kraus_operators_repeater}
    \begin{equation} 
        K_0=\sqrt{1-p_{e}}\mathbb{I}_2
    \end{equation}
     \begin{equation}
        K_1=\sqrt{p_{e}}\sigma_z,
    \end{equation}
\end{subequations}
where $p_e=(1-\sqrt{1-2p})/2$ represents the probability of a phase-flip in the presence of a single node, and $\sigma_z=\begin{pmatrix} 1 & 0 \\ 0 & -1 \end{pmatrix}$. For $n$ channels, the effective dephasing probability between each node is given by $p_e(n)=(1-(1-2p)^{\frac{1}{n}})/2$, which can be used to compute the channel capacity of $(n-1)$ repeaters using $C(n)=1-\text{H}_2(p_e(n))$.

\ozlem{When Alice and Bob each prepare a $\ket{\phi^+}$ Bell state and send one arm to Charlie, the resulting state, as described in equation~\eqref{eq:choi_repeater}, is obtained through the application of the Kraus operators as follows
\begin{align}
    \label{eq:how_to_get_choi_matrix}
    \rho_{\text{AC}}=&\;(\mathbb{I}_2\otimes K_0)\ketbraauto{\phi^+}(\mathbb{I}_2\otimes K_0)^\dagger+ \nonumber \\
    &\;(\mathbb{I}_2\otimes K_1)\ketbraauto{\phi^+}(\mathbb{I}_2\otimes K_1)^\dagger.
\end{align}
This state can be expressed in the computational basis as
\begin{align}
    \label{eq:dephased_state_computational_basis}
    \rho_{\text{AC}}=&\;(1-p_e)\ketbraauto{\phi^+}+p_e\ketbraauto{\phi^-}\nonumber \\
    =&\;\frac{1}{2}\big(\ketbraauto{00}+(1-2p_e)\ket{00}\!\!\bra{11}+\nonumber \\
    &\;(1-2p_e)\ket{11}\!\!\bra{00}+\ketbraauto{11}\big).
\end{align}
}

Therefore, the state received by Charlie from Alice and Bob can be expressed as $\rho_{\text{AC}}^{\otimes 2}$, which takes the following form
\begin{widetext}
   \begin{align}
    \label{eq:tensor_of_two_choi_states}
        \rho_{\mathrm{AC_1C_2B}}=&\;\frac{1}{4}\big(\ketbraauto{0000}+(1-2p_e)\ketbra{0000}{0011}+(1-2p_e)\ketbra{0011}{0000}+\ketbra{0011}{0011}+\nonumber \\
        &\;(1-2p_e)\ketbra{0000}{1100}+(1-2p_e)^2\ketbra{0000}{1111}+(1-2p_e)^2\ketbra{0011}{1100}+(1-2p_e)\ketbra{0011}{1111}+\nonumber \\
        &\;(1-2p_e)\ket{1100}\!\!\bra{0000}+(1-2p_e)^2\ketbra{1100}{0011}+(1-2p_e)^2\ketbra{1111}{0000}+(1-2p_e)\ketbra{1111}{0011}+\nonumber \\
        &\;\ketbraauto{1100}+(1-2p_e)\ketbra{1100}{1111}+(1-2p_e)\ketbra{1111}{1100}+\ketbraauto{1111}\big).
    \end{align} 
\end{widetext}

\ozlem{Charlie then performs a Bell-state measurement, enabling Alice and Bob to share an entangled state, as illustrated in figure~\ref{fig:bsm_charlie}. For the purposes of this proof, assume Charlie uses the $\ket{\phi^+}$ state as the measurement operator. In this case, his measurement is represented by
\begin{widetext}
    \begin{align}
    \label{eq:charlie_measurement}
        \Pi_{\phi^+}=&(\mathbb{I}_2\otimes\ketbraauto{\phi^+}\otimes\mathbb{I}_2)\nonumber \\
        =&\frac{1}{2}\big(\ketbraauto{0000}+\ketbra{0000}{0110}+\ketbra{0110}{0000}+\ketbra{0110}{0110}+\ketbra{1000}{1000}+\ketbra{1000}{1110}+\nonumber \\
        &\ketbra{1110}{1000}+\ketbra{1110}{1110}+\ketbra{0001}{0001}+\ketbra{0001}{0111}+\ketbra{0111}{0001}+\ketbra{0111}{0111}+\nonumber \\
        &\ketbra{1001}{1001}+\ketbra{1001}{1111}+\ketbra{1111}{1001}+\ketbra{1111}{1111}\big).
    \end{align}
\end{widetext}
}
Thus, the state after Charlie's measurement becomes
\begin{widetext}
    \begin{align}
    \label{eq:reduced_state_charlie}
        \rho_{\mathrm{AB}}&=\mathrm{Tr}_{2,3}[\Pi_{\phi^+}\rho_{\mathrm{AC_1C_2B}}\Pi_{\phi^+}^\dagger]\nonumber \\
        &=\frac{1}{8}\big(\ketbraauto{00}+(1-2p_e)^2\ketbra{00}{11}+(1-2p_e)^2\ketbra{11}{00}+\ketbraauto{11}\big),
\end{align}
\end{widetext}
where $\mathrm{Tr}_{2,3}[.]$ represents tracing out the modes 2 and 3 corresponding to $\mathrm{C_1}$ and $\mathrm{C_2}$. Normalising the state by the success probability, $\mathrm{Tr}[\rho_{\mathrm{AB}}]$, gives 
\begin{widetext}
    \begin{align}
    \label{eq:reduced_state_charlie_normalised}
        \Tilde{\rho}_{\mathrm{AB}_{\phi^+}}&=\frac{\rho_{\mathrm{AB}}}{\text{Tr}[\rho_{\mathrm{AB}}]}\nonumber \\
        &=\frac{1}{2}\big(\ketbraauto{00}+(1-2p_e)^2\ketbra{00}{11}+(1-2p_e)^2\ketbra{11}{00}+\ketbraauto{11}\big)\nonumber \\
        &=\frac{1}{2}\big(\ketbraauto{00}+(1-2p)\ketbra{00}{11}+(1-2p)\ketbra{11}{00}+\ketbraauto{11}\big)\nonumber \\
        &=(1-p)\ketbraauto{\phi^+}+p\ketbraauto{\phi^-}.
    \end{align}
\end{widetext}
This demonstrates that the resulting state after the BSM remains identical to the repeaterless case. Each BSM outcome projects the state between Alice and Bob onto a state that experiences the same level of noise as in the repeaterless scenario, with each outcome occurring with a probability of $0.25$ as shown below
\begin{subequations}
\label{eq:other_states}
\begin{equation}
    \Tilde{\rho}_{\mathrm{AB}_{\phi^-}}=(1-p)\ketbraauto{\phi^-}+p\ketbraauto{\phi^+},
\end{equation}
\begin{equation}
    \Tilde{\rho}_{\mathrm{AB}_{\psi^+}}=(1-p)\ketbraauto{\psi^+}+p\ketbraauto{\psi^-},
\end{equation}
\begin{equation}
    \Tilde{\rho}_{\mathrm{AB}_{\psi^-}}=(1-p)\ketbraauto{\psi^-}+p\ketbraauto{\psi^+}.
\end{equation}
\end{subequations}

\ozlem{Therefore the average RCI becomes 
\begin{align}
    \label{eq:average_rci_after_bsm}
    \text{RCI}_\text{BSM}=\;&0.25\times\text{RCI}(\Tilde{\rho}_{\mathrm{AB}_{\phi^+}})+0.25\times\text{RCI}(\Tilde{\rho}_{\mathrm{AB}_{\phi^-}})+\nonumber \\ 
    &0.25\times\text{RCI}(\Tilde{\rho}_{\mathrm{AB}_{\psi^+}})+0.25\times\text{RCI}(\Tilde{\rho}_{\mathrm{AB}_{\psi^-}})\nonumber \\
    &=1-\mathrm{H_2}(p),
\end{align}
which is equal to the capacity of the repeaterless bound~\cite{pirandola2017fundamental}. This highlights the importance of our protocol in achieving the capacity of the reperater bounds of the Pauli dephasing channel.
}

\section{\label{sec:appendix_probability_success}Probability of Success of the Second Stage}
The probability of success of the second round given that the states were successful in the first round is given by
\begin{widetext}
    \begin{equation}
    \begin{aligned}
        P_{s_1s_2}(m) &= \frac{2^{-(m-1)}}{P_{s_1}^m(m)} \sum_{r=0}^{\frac{m-1}{2}} \binom{m}{r} \, \left (1+ \left (1-2 p\right )^{m-2r}\right )^m (1-2 p )^{mr} \quad \quad  \text{for m odd} \, , \\
        P_{s_1s_2}(m) &= \frac{2^{-(m-1)}}{P_{s_1}^m(m)} \left ( \sum_{r=0}^{\frac{m}{2}-1} \binom{m}{r} \, \left (1+ \left (1-2 p\right )^{m-2r}\right )^m (1-2 p )^{mr} + 2^{m-1}(1-2p)^{\frac{m^2}{2}}\binom{m}{m/2}  \right ) \quad \quad  \text{for m even} \, , 
    \end{aligned}
\end{equation}
\end{widetext}
where $r$ represents half the number of errors of the first stage.

\section{\label{sec:appendix_eigenvalues_second_stage}Eigenvalues of the Second Stage}
\ZJ{As mentioned in Sec.~\ref{sec:second_stage}, in order to saturate the channel capacity, we need to keep all the possible outcomes. Accordingly, we provide the eigenvalues of the remaining conditional states here. The eigenvalues of the successful state in the second round, given it failed the first round, $\lambda_{f_1s_2}(j)$, follow the same formula as equation~\eqref{eq:eigenvalues_second_success_success} with $j = m, m+2, \ldots, m(m-1)$ for even $m$ and $j = m, m+2, \ldots, m(m-1)+1$ for odd $m$. Similar to equation~\eqref{eq:eigenvalue_second_success_success_normalised}, these eigenvalues need to be normalised. The probability of success in the second round, $P_{f_1s_2}(m)$, given it failed in the first round, is computed using equation~\eqref{eq:ps_of_second_round} with $j$ values adjusted accordingly for the failed states and $\Tilde{\lambda}_{s_1s_2}(j)$ is divided by $P_{f_1}^m(m)$ where $P_{f_1}^m(m)$ is obtained from equation~\eqref{eq:prob_failure_stage1}. Therefore, the eigenvalues of the successful state of the second round given it fails the first round is given by $\lambda_{f_1s_2}(j)=\Tilde{\lambda}_{s_1s_2}(j)/(P_{f_1}^m(m)P_{f_1s_2}(m))$.}

\ZJ{The last set of eigenvalues we need to compute are for the states that failed in both the first and second rounds. These are derived from the relationship between the failed and successful states of the second round and the failed state from the first round, as shown below
\begin{equation}
    \rho^{\otimes (m-1)}_{\mathrm{AB}_{f_1}}=P_{f_1s_2}(m)\times\rho_{\mathrm{AB}_{f_1s_2}}+P_{f_1f_2}(m)\times\rho_{\mathrm{AB}_{f_1f_2}},
\label{eq:link_between_first_and_second_stage_given_first_failed}
\end{equation}
Here, $\rho_{\mathrm{AB}_{f_1}}$ represents the failed state from the first stage of the protocol. $P_{f_1s_2}(m)$ and $P_{f_1f_2}(m)$ are the probabilities of success and failure, respectively, in the second round given the state failed in the first round. Note that $P_{f_1f_2}(m)$ is equal to $1 - P_{f_1s_2}(m)$. $\rho_{\mathrm{AB}_{f_1s_2}}$ and $\rho_{\mathrm{AB}_{f_1f_2}}$ are the conditional states of the second round that succeeded and failed, respectively, given they failed in the first round. Therefore the eigenvalues of the failed state of the second round given it also failed the first stage is given by
\begin{multline}
    \lambda_{f_1f_2}(j_1,j_2)=\frac{1}{P_{f_1f_2}(m)}\bigg[\frac{(1-p)^{m(m-1)-j_1}p^{j_1}}{P_{f_1}^{m-1}(m)}\\-\frac{(1-p)^{m^2-j_2}p^{j_2}}{P_{f_1}^m(m)}\bigg],
    \label{eq:eigenvalue_second_round_fail_fail}
\end{multline}
where $j_1=m-1,m+1,\cdots,(m-1)^2$ and $j_2=j_1+1,j_1+3,\cdots,j_1+m-1$ for even $m$ and $j_1=m-1,m+1,\cdots,m(m-1)$ and $j_2=j_1+1,j_1+3,\cdots,j_1+m$ for odd $m$.
}

\section{\label{sec:appendix_cap_first}Proof of the Capacity for the First Round of the Protocol}
As stated in equation~\eqref{eq:rci_first_stage_relation} when we add the successful and failed states up after the first round, they give the state of $m-1$ copies of Bell pairs as in average nothing changes but the state conditioned on a successful outcome individually purifies the Bell pairs. Note that the eigenvalues of the successful state is $\frac{(1-p)^{m-j_1}p^{j_1}}{P_{s_1}(m)}$ with a multiplicity of $\binom{m}{j_1}$ while the eigenvalues of the failed state is $\frac{(1-p)^{m-(j_1+1)}p^{j_1+1}}{P_{f_1}(m)}$ with a multiplicity of $\binom{m}{j_1+1}$ with $j_1=0,2,\cdots,m$ if $m$ is even. Therefore, when we add the successful and failed states up, the eigenvalues simply become $(1-p)^{(m-1)}p^{j_1}$ with $j_1=0,1,\cdots,m-1$ with each eigenvalue having a multiplicity of $\binom{m-1}{j_1}$. Then the entropy of the combined state in the RHS of equation~\eqref{eq:rci_first_stage_relation} becomes
\begin{widetext}
    \begin{align}
    \label{eq:rhs_first_stage_proof}
    \mathrm{RHS}&=\frac{1}{m}\big(S(P_{s_1}(m)\rho_{\mathrm{A}_{s_1}}+P_{f_1}(m)\rho_{\mathrm{A}_{f_1}})-S(P_{s_1}(m)\rho_{\mathrm{AB}_{s_1}}+P_{f_1}(m)\rho_{\mathrm{AB}_{f_1}})\big)\nonumber \\
    &=\frac{1}{m}\mathrm{log_2}(2^{m-1})+\frac{1}{m}\sum_{j_1=0}^{m-1}\binom{m-1}{j_1}(1-p)^{(m-1)-j_1}p^{j_1}\mathrm{log_2}\big((1-p)^{(m-1)-j_1}p^{j_1}\big) \nonumber \\
    &=\frac{m-1}{m}+\frac{(1-p)^{m-1}}{m}\bigg[\sum_{j_1=0}^{m-1}\binom{m-1}{j_1}(1-p)^{(m-1)-j_1}p^{j_1} \big((m-1-j_1)\mathrm{log_2}(1-p)+j_1\mathrm{log_2}(p)\big)\bigg] \nonumber \\
    &=\frac{m-1}{m}+\frac{(1-p)^{m-1}}{m}\bigg[\sum_{j_1=0}^{m-1}\binom{m-1}{j_1}\bigg(\frac{p}{1-p}\bigg)^{j_1}(m-1)\mathrm{log_2}(1-p)+\sum_{j_1=0}^{m-1}\binom{m-1}{j_1}\bigg(\frac{p}{1-p}\bigg)^{j_1}\nonumber \\
    &j_1\big(\mathrm{log_2}(p)-\mathrm{log_2}(1-p)\big)\bigg].
\end{align}
\end{widetext}
Note that $S(P_{s_1}(m)\rho_{\mathrm{A}_{s_1}}+P_{f_1}(m)\rho_{\mathrm{A}_{f_1}})=\mathrm{log_2}(2^{m-1})$ and using $\sum_{j=0}^n\binom{n}{k} x^j=(1+x)^{n}$ and $\sum_{j=0}^n\binom{n}{k}j x^j=n x(1+x)^{n-1}$, the expression in equation~\eqref{eq:rhs_first_stage_proof} can be further simplified to
\begin{widetext}
\begin{align}
\label{eq:rhs_firs_stage_proof_complete}
    \mathrm{RHS}&=\frac{m-1}{m}+\frac{(1-p)^{m-1}(m-1)}{m}\bigg[\bigg(1+\frac{p}{1-p}\bigg)^{m-1}\mathrm{log_2}(1-p)-\bigg(\frac{p}{1-p}\bigg)\bigg(1+\frac{p}{1-p}\bigg)^{m-2}\mathrm{log_2}(1-p)\nonumber \\
    &+\bigg(\frac{p}{1-p}\bigg)\bigg(1+\frac{p}{1-p}\bigg)^{m-2}\mathrm{log_2}(p) \nonumber
    \bigg] \nonumber \\
    &=\frac{m-1}{m}+\frac{(1-p)^{m-1}(m-1)}{m}\bigg(1+\frac{p}{1-p}\bigg)^{m-2}\bigg[\mathrm{log_2}(1-p)+\frac{p}{1-p}\mathrm{log_2}(p)\bigg] \nonumber \\
    &=\frac{m-1}{m}\big[1-\big(-(1-p)\mathrm{log_2}(1-p)-(p)\mathrm{log_2}(p)\big)\big] \nonumber \nonumber \\
    &=\frac{m-1}{m}\big(1-\mathrm{H}_2(p)\big) \nonumber \\
    &=\frac{m-1}{m}\,C,
\end{align}
\end{widetext}
as $1-\mathrm{H_2}(p)$ is equal to the capacity of the dephasing channel. Additionally, the LHS of equation~\eqref{eq:rci_first_stage_relation} is already proven in Sec.~\ref{sec:achieving_capacity}.

Alternatively, we can use the following proof utilising the properties of von Neumann entropy which is also helpful for the proof of the second stage of the protocol. Recall equation~\eqref{eq:average_rci_stage1} which presents the average RCI after the first round of the protocol. This equation can be expanded as
\begin{align}
\label{eq:average_rci_stage1_expanded}
    \mathrm{RCI_1}&=\frac{1}{m}\big[P_{s_1}(m)\,S(\rho_{\mathrm{A}_{s_1}})+P_{f_1}(m)\,S(\rho_{\mathrm{A}_{f_1}})\nonumber \\
    &-P_{s_1}(m)\,S(\rho_{\mathrm{AB}_{s_1}})-P_{f_1}(m)\,S(\rho_{\mathrm{AB}_{f_1}})\big].
\end{align}
As both $S(\rho_\mathrm{A_{s_1}})$ and $S(\rho_\mathrm{A_{f_1}})$ are equal to the dimensions of the state which is $\log_2(2^{m-1})$. Therefore, $P_{s_1}(m)\,S(\rho_{\mathrm{A}_{s_1}})+P_{f_1}(m)\,S(\rho_{\mathrm{A}_{f_1}})=\log_2(2^{m-1})\times(P_{s_1}(m)+P_{f_1}(m))=(m-1)$ as $P_{s_1}(m)+P_{f_1}(m)=1$. Therefore,
\begin{align}
\label{eq:alternative_proof_stage1}
 &\mathrm{RCI_1}\!=\!\frac{1}{m}\big[(m-1)\!-\!P_{s_1}(m)S(\rho_{\mathrm{AB}_{s_1}})\!-\!P_{f_1}(m)S(\rho_{\mathrm{AB}_{f_1}})\big].
\end{align}
Using concavity of von Neumann entropy where $S(\sum_i\lambda_i\rho_i)\geq\sum_i\lambda_i\,S(\rho_i)$ for $\lambda_i\geq0$ and $\sum_i\lambda_i=1$,
\begin{multline}
\label{eq:entropy_relation_stage1_alternative_proof}
     P_{s_1}(m)\,S(\rho_{\mathrm{AB}_{s_1}})+P_{f_1}(m)\,S(\rho_{\mathrm{AB}_{f_1}})\leq\,\\ S(P_{s_1}(m)\times\rho_{\mathrm{AB}_{s_1}}+P_{f_1}(m)\times\rho_{\mathrm{AB}_{f_1}}).
\end{multline}
Therefore by substituting the above relation into equation~\eqref{eq:alternative_proof_stage1} we can lower bound the actual RCI of this protocol by
\begin{widetext}
\begin{align}
\label{eq:alternative_proof_stage1_complete}
    &\frac{1}{m}\big[(m-1)-S( P_{s_1}(m)\times\rho_{\mathrm{AB}_{s_1}}+P_{f_1}(m)\times\rho_{\mathrm{AB}_{f_1}})\big]\leq\frac{1}{m}\big[(m-1)-P_{s_1}(m)\,S(\rho_{\mathrm{AB}_{s_1}})-P_{f_1}(m)\,S(\rho_{\mathrm{AB}_{f_1}})\big]\leq\,C \nonumber \\
    &\frac{m-1}{m}\,C\leq\frac{1}{m}\big[(m-1)-P_{s_1}(m)\,S(\rho_{\mathrm{AB}_{s_1}})-P_{f_1}(m)\,S(\rho_{\mathrm{AB}_{f_1}})\big]\leq\,C \nonumber \\
    &\frac{m-1}{m}\,C\leq\mathrm{RCI_1}\leq\,C
\end{align}
\end{widetext}
The proof of the lower bound in equation~\eqref{eq:alternative_proof_stage1_complete}, follows the same steps as those in equation~\eqref{eq:rhs_first_stage_proof}. Therefore, refer to equation~\eqref{eq:rhs_first_stage_proof} for the expansion of $S( P_{s_1}(m)\times\rho_{\mathrm{AB}_{s_1}}+P_{f_1}(m)\times\rho_{\mathrm{AB}_{f_1}})$. Notably, equation~\eqref{eq:alternative_proof_stage1_complete} becomes the capacity, $C$ as $\lim_{m\to\infty}\frac{m-1}{m}\,C=C$. 
\\
\section{\label{sec:appendix_cap_second}Proof of the Capacity for the Second Round of the Protocol}
Recall that equation~\eqref{eq:average_rci_stage2} represents the average RCI for the second round of the protocol. However, unlike in the first stage, the eigenvalue multiplicities are not available in closed form, making it challenging to directly prove equation~\eqref{eq:average_rci_stage2}. To address this, we substitute $P_{s_1s_2}(m)\times\mathrm{RCI}(\rho_{\mathrm{AB}_{s_1s_2}}) +P_{s_1f_2}(m)\times\mathrm{RCI}(\rho_{\mathrm{AB}_{s_1f_2}})$ with $\mathrm{RCI}(\rho_{\mathrm{AB}_{s_1}}^{\otimes (m-1)})$, as averaging over the conditional states, whether successful or failed in the second round, given their success in the first, yields $m-1$ copies of the successful state from the first round. Similarly, we replace $P_{f_1s_2}(m)\times\mathrm{RCI}(\rho_{\mathrm{AB}_{f_1s_2}}) +P_{f_1f_2}(m)\times\mathrm{RCI}(\rho_{\mathrm{AB}_{f_1f_2}})$ with $\mathrm{RCI}(\rho_{\mathrm{AB}_{f_1}}^{\otimes (m-1)})$, since these terms also sum to $m-1$ copies of the failed state from the first round, thereby conserving the reverse coherent information. However, computing the RCI of these states underestimates the key rates. Therefore, we demonstrate that equation~\eqref{eq:average_rci_stage2} can be lower bounded by
\begin{widetext}
    \begin{align}
    \label{eq:ave_rci_relation_second_stage}
    &\frac{1}{m^2}\big[P_{s_1}(m)\times\mathrm{RCI}(\rho_{\mathrm{AB}_{s_1}}^{\otimes (m-1)})+P_{f_1}(m)\times\mathrm{RCI}(\rho_{\mathrm{AB}_{f_1}}^{\otimes (m-1)})\big]\leq\,\mathrm{RCI_2}\leq\,C \nonumber \\
    &\frac{1}{m^2}\big[P_{s_1}(m)\,S(\rho_{\mathrm{A}_{s_1}}^{\otimes(m-1)})+P_{f_1}(m)\,S(\rho_{\mathrm{A}_{f_1}}^{\otimes(m-1)})-P_{s_1}(m)\,S(\rho_{\mathrm{AB}_{s_1}}^{\otimes (m-1)})-P_{f_1}(m)\,S(\rho_{\mathrm{AB}_{f_1}}^{\otimes (m-1)})\big]\leq\,\mathrm{RCI_2}\leq\,C \nonumber \\
    &\frac{1}{m^2}\big[(m-1)^2-\big(P_{s_1}(m)\,S(\rho_{\mathrm{AB}_{s_1}}^{\otimes (m-1)})+P_{f_1}(m)\,S(\rho_{\mathrm{AB}_{s_1}}^{\otimes (m-1)})\big)\big]\leq\,\mathrm{RCI_2}\leq\,C,
\end{align}
\end{widetext}
where both $S(\rho_{\mathrm{A}_{s_1}}^{\otimes(m-1)})$ and $S(\rho_{\mathrm{A}_{f_1}}^{\otimes(m-1)})$ are just equal to the dimensions of the state which is $\log_2(2^{(m-1)^2})$. Therefore, $P_{s_1}(m)\,S(\rho_{\mathrm{A}_{s_1}}^{\otimes(m-1)})+P_{f_1}(m)\,S(\rho_{\mathrm{A}_{f_1}}^{\otimes(m-1)})=\log_2(2^{(m-1)^2})\times(P_{s_1}(m)+P_{f_1}(m))=(m-1)^2$ as $P_{s_1}(m)+P_{f_1}(m)=1$. As both $\rho_{\mathrm{AB}_s1}^{\otimes (m-1)}$ and $\rho_{\mathrm{AB}_f1}^{\otimes (m-1)}$ have no entanglement between each state, we can use $S(\rho^{\otimes n})=n\,S(\rho)$, the relation becomes 
\begin{widetext}
    \begin{align}
    \label{eq:ave_rci_relation_second_stage_continued}
    &\frac{1}{m^2}\big[(m-1)^2-(m-1)\big(P_{s_1}(m)\,S(\rho_{\mathrm{AB}_{s_1}})+P_{f_1}(m)\,S(\rho_{\mathrm{AB}_{s_1}})\big)\big]\leq\,\mathrm{RCI_2}\leq\,C.
\end{align}
\end{widetext}

Using the concavity of von Neumann entropy, the LHS of equation~\eqref{eq:ave_rci_relation_second_stage_continued} can be lower bounded by
\\
\begin{widetext}
    \begin{align}
    &\frac{1}{m^2}\big[(m-1)^2\!-\!(m-1)S\big(P_{s_1}(m)\rho_{\mathrm{AB}_{s_1}}\!+\!P_{f_1}(m)\rho_{\mathrm{AB}_{s_1}}\big)\big]
    \!\leq\!\frac{1}{m^2}\big[(m-1)^2\!-\!(m-1)\big(P_{s_1}(m)S(\rho_{\mathrm{AB}_{s_1}})\!+\!P_{f_1}(m)S(\rho_{\mathrm{AB}_{s_1}})\big)\big] \nonumber \\
    &\frac{1}{m^2}\big[(m-1)^2-(m-1)^2\,\mathrm{H_2}(p)\big]\leq\frac{1}{m^2}\big[(m-1)^2-(m-1)\big(P_{s_1}(m)S(\rho_{\mathrm{AB}_{s_1}})+P_{f_1}(m)S(\rho_{\mathrm{AB}_{s_1}})\big)\big] \nonumber \\
    &\bigg(\frac{m-1}{m}\bigg)^2\big[1-\,\mathrm{H_2}(p)\big]\leq\frac{1}{m^2}\big[(m-1)^2-(m-1)\big(P_{s_1}(m)S(\rho_{\mathrm{AB}_{s_1}})+P_{f_1}(m)S(\rho_{\mathrm{AB}_{s_1}})\big)\big]\nonumber \\
    &\bigg(\frac{m-1}{m}\bigg)^2\,C\leq\frac{1}{m^2}\big[(m-1)^2-(m-1)\big(P_{s_1}(m)S(\rho_{\mathrm{AB}_{s_1}})+P_{f_1}(m)S(\rho_{\mathrm{AB}_{s_1}})\big)\big].
    \end{align}
\end{widetext}
Therefore, the RCI of the second round can be expressed as
\begin{equation}
\label{eq:rci_second_stage_versus_capacity}
    \bigg(\frac{m-1}{m}\bigg)^2\,C\leq\,\mathrm{RCI_2}\leq\,C.
\end{equation}

\section{\label{sec:appendix_any_round}Key Rate of the Protocol in Any Round}
As the RCI of the first round is $\frac{m-1}{m}\,C$ and the RCI of the second stage is lower bounded by equation~\eqref{eq:rci_second_stage_versus_capacity}, the relationship between the RCI of any stage of the protocol versus capacity can be gives as
\begin{equation}
\label{eq:rci_of_any_round}
    \bigg(\frac{m-1}{m}\bigg)^n\,C\leq\,\mathrm{RCI}_n\leq\,C,
\end{equation}
where $n$ denotes the number of rounds. As mentioned in the main text, if $n=m$, as $m\rightarrow\infty$, the limit approaches $C/e$, which would cause the lower bound to fall below the capacity. However, the effective dephasing decreases with each round, resulting in higher distillable entanglement. As long as $n\ll m$, the protocol is able to reach the capacity of the dephasing channel. 
%However, if the values of $n$ and $m$ are comparable, then the protocol will be
%\begin{equation}
%    \lim_{m\to\infty}=\bigg(\frac{m-1}{m}\bigg)^{m/n}\,C=\frac{1}{e^{1/n}}\,C.
%\end{equation}

\section{\label{sec:appendix_fidelity}Proof of Fidelity}
Following the first round of the protocol, the reduced states in the alternative protocol have the same form as in the original protocol. However, in the second round, the form of the reduced state in the original protocol deviates from Eq~\eqref{eq:reduced_density_matrix}. Nonetheless, since the fidelity of our actual protocol is higher than that of the alternative protocol, we use the alternative protocol as a lower bound to demonstrate that fidelity improves with each round, while the dephasing probability decreases.

Since the fidelity is defined as $F=1-p$ for the dephasing channel, demonstrating that $p$ progressively decreases implies that the fidelity increases with each iteration. For this reason, the relationship between the dephasing probability for each round of protocol can be expressed as 
\begin{equation}
\label{eq:iterative_map_probability}
    p_{n+1}=\Tilde{p}(p_n),
\end{equation}
where the new probability of dephasing depends on the dephasing probability of the previous one. Therefore in reach round,
\begin{subequations}
\label{eq:iterative_map_equations}
    \begin{equation}
        p_1=\Tilde{p}(p_0),
    \end{equation}
    \begin{equation}
        p_2=\Tilde{p}(p_1)=\Tilde{p}(\Tilde{p}(p_0)),
    \end{equation}
    \begin{equation}
        p_3=\Tilde{p}(p_2)=\Tilde{p}(\Tilde{p}(\Tilde{p}(p_0))),
    \end{equation}
\end{subequations}
where $\Tilde{p}(p)$ is the new dephasing probability as a function of the previous dephasing probability which is given in equation~\eqref{eq:new_dephasing_probability}. Based on this, let us prove that
\begin{equation}
    \label{eq:prob_eqaution_to_prove_one}
    p_1=\Tilde{p}(p_0) \leq p_0=p.
\end{equation}
\ZJ{Recall that for the dephasing channel, we have the constraint $0 \leq p_0 \leq \frac{1}{2}$. Using this range, we can derive the following sequence of inequalities to show that $p_1 \leq p_0$.}
\ZJ{\begin{widetext}
    \begin{align}
        \label{eq:inequality_generic}
        0 \leq p_0 &\leq\frac{1}{2} \nonumber \\
        \Leftrightarrow 0\leq 2p_0 &\leq 1 \nonumber \\
        \Leftrightarrow 0\leq 1-2p_0 &\leq 1 \nonumber \\
        \Leftrightarrow 0\leq (1-2p_0)^2 &\leq 1 \nonumber \\
        \Leftrightarrow 0\leq (1-2p_0)^2(1-2p_0)^m &\leq (1-2p_0)^m \nonumber \\
        \Leftrightarrow (1-2p_0)^m &\leq (1-2p_0)^{m-2} \nonumber \\
        \Leftrightarrow 1+(1-2p_0)^m &\leq 1+(1-2p_0)^{m-2} \nonumber \\
        \Leftrightarrow (1-2p_0)(1+(1-2p_0)^m) &\leq (1-2p_0)(1+(1-2p_0)^{m-2}) \nonumber \\
        \Leftrightarrow 1-2p_0 &\leq \frac{(1-2p_0)+(1-2p_0)^{m-1}}{(1+(1-2p_0)^m)} \nonumber \\
        \Leftrightarrow 1-\frac{(1-2p_0)+(1-2p_0)^{m-1}}{(1+(1-2p_0)^m)} &\leq 2p_0 \nonumber \\
        \Leftrightarrow p_1=\frac{1}{2}\bigg(1-\frac{(1-2p_0)+(1-2p_0)^{m-1}}{(1+(1-2p_0)^m)}\bigg) &\leq p_0.
    \end{align}
\end{widetext}
}
\ZJ{This final expression in equation~\eqref{eq:inequality_generic} on the left-hand side matches equation~\eqref{eq:new_dephasing_probability} as a function of any Bell pair, $m$, which confirms that
\begin{equation}
    \label{eq:p1_less_than_p0}
    p_1\leq p_0.
\end{equation}
}
%\ZJ{Using equation~\eqref{eq:new_dephasing_probability}, in the following proof we show that $p_1\leq p_0$, where the RHS of the inequality below is $p_1$.}
%\begin{equation}
%    \label{eq:prob_eqaution_to_prove}
%    \frac{1}{2}\bigg(1-\frac{(1-2p_0)+(1-2p_0)^{m-1}}{(1+(1-2p_0)^m)}\bigg) \leq p_0,
%\end{equation}
%and substituting $p_0=(1-2z)/2$ into equation~\eqref{eq:prob_eqaution_to_prove}
%\begin{align}
%    \label{eq:proof_of_prob_first_stage}
%    \frac{(-1+2z)(-2z+2^m z^m)}{4z(1+2^m z^m)} &\leq \frac{1-2z}{2} \nonumber \\
%    \Leftrightarrow\frac{2z(1-2z)(1-2^{m-1}z^{m-1})}{4z(1+2^m z^m)} &\leq \frac{1-2z}{2} \nonumber \\
%    \Leftrightarrow\frac{(1-2^{m-1}z^{m-1})}{(1+2^m z^m)} &\leq 1 \nonumber \\
%    \Leftrightarrow1-2^{m-1}z^{m-1} &\leq 1+2^m z^m \nonumber \\ 
%    \Leftrightarrow2^m z^m &\leq 2^{m-1}z^{m-1}\nonumber \\
%    \Leftrightarrow2z &\leq 1 \nonumber \\
%    \Leftrightarrow1-2p_0 &\leq 1,
%\end{align}
%\begin{equation}
%\label{eq:end_points_prob_first_stage}
%    \begin{cases}
%    1-2p=0\leq 1, & p=1/2 \\
%    1-2p=1\leq 1, & p=0,
%\end{cases}
%\end{equation}
%\ZJ{therefore, when $p_0\leq1/2$,
%\begin{equation}
%    \label{eq:p1_less_than_p0}
%    \implies p_1\leq p_0.
%\end{equation}}
\ZJ{Note that the proof of $p_{n+1} \leq p_n$ for each round of the protocol follows from equation~\eqref{eq:inequality_generic}. For example, in the second round, each occurrence of $p_0$ is replaced by $p_1$, leading to $p_2 \leq p_1 \leq p_0$, which in turn implies $F_2 \geq F_1 \geq F_0$. To show that $\lim_{n \rightarrow \infty} \tilde{p}(p_n) = 0$ and $\lim_{n \rightarrow \infty} F = 1$, we apply fixed-point and stability analysis techniques from non-linear dynamics~\cite{schuster2006deterministic} to the iterative map function $\tilde{p}(p_n)$. We begin by computing the fixed points of $\tilde{p}(p_n) = p_n$,
\begin{align}
    \label{eq:fixed_points}
    \frac{1}{2}\bigg(1-\frac{(1-2p_n)+(1-2p_n)^{m-1}}{(1+(1-2p_n)^m)}\bigg)&=p_n \nonumber \\
    p_n&=0,1.
\end{align}
We disregard $p_n = 1$ since, for the dephasing channel, $p_n$ is always constrained to the range $0 \leq p_n \leq 1/2$. Next, we compute the derivative of $\tilde{p}(p_n)$ at the fixed point $p_n = 0$, given by $\left. \frac{\text{d}\tilde{p}}{\text{d}p_n} \right|_{p_n = 0}$. In this case,
\begin{equation}
    \label{eq:derivative_of_p}
    \frac{\text{d}\Tilde{p}}{\text{d}p_n}\!=\!\frac{1\!-\!(1\!-\!2p_n)^{2m}\!+\!4p_n(p_n\!-\!1)(1\!-\!(m\!-\!1)(1\!-\!2p_n)^m)}{(1\!+\!(1\!-\!2p_n)^m)^2(1\!-\!2p_n)^2},
\end{equation}
and substituting $p_n=0$ into the expression above, we obtain
\begin{equation}
    \label{eq:derivative_at_fixed_point}
    \left| \left. \frac{\mathrm{d} \tilde{p}}{\mathrm{d} p_n} \right|_{p_n = 0} \right|=0.
\end{equation}
Note that if $\left| \left. \frac{\mathrm{d} \tilde{p}}{\mathrm{d} p_n} \right|_{p_n = 0} \right| < 1$, the fixed point $p_n = 0$ is considered stable~\cite{schuster2006deterministic}, meaning the iterative map converges to this fixed point. In our case, since the derivative at this fixed point is also zero, it is classified as super-stable~\cite{schuster2006deterministic}. This implies that $p_n$ converges to zero very rapidly, leading to the fidelity approaching one as the number of rounds increases.
}
%When the protocol is repeated, the relationship between the dephasing probabilities of the second and first rounds become
%\begin{align}
%\label{eq:proof_of_prob_second_stage}
%    p_1\!&\!\geq p_2 \nonumber \\
%    p_1&\!\geq\!\frac{1}{2}\!\bigg(\!1\!-\!\frac{(1-2p_1)^{m-1}\!+\!(1-2p_1)}{1+(1-2p_1)^m}\!\bigg) \nonumber \\
%    1-2p_1&\!\leq\!\frac{(1-2p_1)^{m-1}+(1-2p_1)}{1+(1-2p_1)^m} \nonumber \\
%    (1-2p_1)+(1-2p_1)^m &\!\leq (1-2p_1)+(1-2p_1)^{m-1}\nonumber \\
%    (1-2p_1)^2 &\!\leq1.
%\end{align}
%Note that, when $p=0, p_1=0$ and $p=1/2, p_1=1/2$ for $m\geq2$. Therefore,
%\begin{equation}
%\label{eq:end_points_prob_second_stage}
%    \begin{cases}
%    (1-2p_1)^2=0\leq 1, & p=0 \\
%    (1-2p_1)^2=1\leq 1, & p=1/2.
%\end{cases}
%\end{equation}
%This demonstrates that $p_2\leq p_1\leq p_0$, which in turn implies that $F_2\geq F_1\geq F_0$. Since this serves as a lower bound to the actual protocol, it also confirms that the actual protocol achieves purification.

%Can we prove the following?
%\begin{widetext}
%\begin{equation}
%RCI_2=\frac{(m-1)}{m}[1-P_{s_1}(m)\times H_2(p_s)-%P_{f_1}(m)\times H_2(p_f)],    
%\end{equation}  
%\end{widetext}

%where $p_s=\frac{1}{2}\bigg(1-\frac{(1-2p)+(1-2p)^{m-1}}{(1+(1-2p)^m)}\bigg)$ and $p_f=\frac{1}{2}\bigg(1-\frac{(1-2p)-(1-2p)^{m-1}}{(1-(1-2p)^m)}\bigg)$

\section{\label{sec:appendix_cost_of_purification}Cost of Purification Estimation}
\ozlemREV{In this section, we show how the cost of purification is estimated. For small dephasing probability $p$, we expand the recursion relation around $p = 0$ and keep only the leading terms, which provide a good approximation in this regime. For the first three rounds, and for any $m$ with small $p$, the Taylor expansion takes the form
\begin{subequations}
\label{eq:taylor_fidelity}
    \begin{equation}
        \Tilde{p}_1=(m-1)p^2+\mathcal{O}(p^3),
    \end{equation}
    \begin{equation}
        \Tilde{p}_2=(m-1)^3p^4+\mathcal{O}(p^5),
    \end{equation}
    \begin{equation}
        \Tilde{p}_2=(m-1)^7p^8+\mathcal{O}(p^9),
    \end{equation}
\end{subequations}
which leads to Eq.~\eqref{eq:new_dephasing_prob_round_n} for any purification round $n$. Since the fidelity satisfies $F = 1 - \Tilde{p}_n$ and we set this equal to a target value $\delta$, we can derive Eq.~\eqref{eq:round_vs_delta_fidelity} by neglecting higher-order terms from
\begin{align}
    \label{eq:derive_n_values}
    p^{2^n}(m-1)^{2^n-1}&=\delta\nonumber \\
    p^{2^n}(m-1)^{2^n}&=\delta(m-1)\nonumber \\
    2^n\log\big(p(m-1)\big)&=\log\big(\delta(m-1)\big)\nonumber \\
    n&=\log_2\bigg(\frac{\log\big(\delta(m-1)\big)}{\log\big(p(m-1)\big)}\bigg).
\end{align}
}

\FloatBarrier

\bibliography{references}% Produces the bibliography via BibTeX.

@article{bennett1996purification,
  title={Purification of noisy entanglement and faithful teleportation via noisy channels},
  author={Bennett, Charles H and Brassard, Gilles and Popescu, Sandu and Schumacher, Benjamin and Smolin, John A and Wootters, William K},
  journal={Phys. Rev. Lett.},
  volume={76},
  number={5},
  pages={722},
  year={1996},
  publisher={APS}
}

@article{bennett1996mixed,
  title={Mixed-state entanglement and quantum error correction},
  author={Bennett, Charles H and DiVincenzo, David P and Smolin, John A and Wootters, William K},
  journal={Phys. Rev. A},
  volume={54},
  number={5},
  pages={3824},
  year={1996},
  publisher={APS}
}

@article{deutsch1996quantum,
  title={Quantum privacy amplification and the security of quantum cryptography over noisy channels},
  author={Deutsch, David and Ekert, Artur and Jozsa, Richard and Macchiavello, Chiara and Popescu, Sandu and Sanpera, Anna},
  journal={Phys. Rev. Lett.},
  volume={77},
  number={13},
  pages={2818},
  year={1996},
  publisher={APS}
}

@article{horodecki1997inseparable,
  title={Inseparable two spin-1 2 density matrices can be distilled to a singlet form},
  author={Horodecki, Micha{\l} and Horodecki, Pawe{\l} and Horodecki, Ryszard},
  journal={Phys. Rev. Lett.},
  volume={78},
  number={4},
  pages={574},
  year={1997},
  publisher={APS}
}

@book{nielsen2010quantum,
  title={Quantum computation and quantum information},
  author={Nielsen, Michael A and Chuang, Isaac L},
  year={2010},
  publisher={Cambridge University Press}
}

@article{pirandola2017fundamental,
  title={Fundamental limits of repeaterless quantum communications},
  author={Pirandola, Stefano and Laurenza, Riccardo and Ottaviani, Carlo and Banchi, Leonardo},
  journal={Nat. Commun.},
  volume={8},
  number={1},
  pages={15043},
  year={2017},
  publisher={Nature Publishing Group UK London}
}

@article{pirandola2009direct,
  title={Direct and reverse secret-key capacities of a quantum channel},
  author={Pirandola, Stefano and Garc{\'\i}a-Patr{\'o}n, Raul and Braunstein, Samuel L and Lloyd, Seth},
  journal={Phys. Rev. Lett.},
  volume={102},
  number={5},
  pages={050503},
  year={2009},
  publisher={APS}
}

@article{pirandola2019end,
  title={End-to-end capacities of a quantum communication network},
  author={Pirandola, Stefano},
  journal={Commun. Phys.},
  volume={2},
  number={1},
  pages={51},
  year={2019},
  publisher={Nature Publishing Group UK London}
}

@article{briegel1998quantum,
  title={Quantum repeaters: the role of imperfect local operations in quantum communication},
  author={Briegel, H-J and D{\"u}r, Wolfgang and Cirac, Juan I and Zoller, Peter},
  journal={Phys. Rev. Lett.},
  volume={81},
  number={26},
  pages={5932},
  year={1998},
  publisher={APS}
}

@article{dur1999quantum,
  title={Quantum repeaters based on entanglement purification},
  author={D{\"u}r, Wolfgang and Briegel, H-J and Cirac, Juan Ignacio and Zoller, Peter},
  journal={Phys. Rev. A},
  volume={59},
  number={1},
  pages={169},
  year={1999},
  publisher={APS}
}

@article{munro2015inside,
  title={Inside quantum repeaters},
  author={Munro, William J and Azuma, Koji and Tamaki, Kiyoshi and Nemoto, Kae},
  journal={IEEE J. Sel. Top. Quantum Electron.},
  volume={21},
  number={3},
  pages={78--90},
  year={2015},
  publisher={IEEE}
}

@article{azuma2022quantum,
  title={Quantum repeaters: From quantum networks to the quantum internet},
  author={Azuma, Koji and Economou, Sophia E and Elkouss, David and Hilaire, Paul and Jiang, Liang and Lo, Hoi-Kwong and Tzitrin, Ilan},
  journal={Rev. Mod. Phys.},
  volume={95},
  number={4},
  pages={045006},
  year={2023},
  publisher={APS}
}

@article{muralidharan2016optimal,
  title={Optimal architectures for long distance quantum communication},
  author={Muralidharan, Sreraman and Li, Linshu and Kim, Jungsang and L{\"u}tkenhaus, Norbert and Lukin, Mikhail D and Jiang, Liang},
  journal={Sci. Rep.},
  volume={6},
  number={1},
  pages={20463},
  year={2016},
  publisher={Nature Publishing Group UK London}
}

@article{lucamarini2018overcoming,
  title={Overcoming the rate--distance limit of quantum key distribution without quantum repeaters},
  author={Lucamarini, Marco and Yuan, Zhiliang L and Dynes, James F and Shields, Andrew J},
  journal={Nature},
  volume={557},
  number={7705},
  pages={400--403},
  year={2018},
  publisher={Nature Publishing Group UK London}
}

@article{wang2018twin,
  title={Twin-field quantum key distribution with large misalignment error},
  author={Wang, Xiang-Bin and Yu, Zong-Wen and Hu, Xiao-Long},
  journal={Phys. Rev. A},
  volume={98},
  number={6},
  pages={062323},
  year={2018},
  publisher={APS}
}

@article{cui2019twin,
  title={Twin-field quantum key distribution without phase postselection},
  author={Cui, Chaohan and Yin, Zhen-Qiang and Wang, Rong and Chen, Wei and Wang, Shuang and Guo, Guang-Can and Han, Zheng-Fu},
  journal={Phys. Rev. Appl.},
  volume={11},
  number={3},
  pages={034053},
  year={2019},
  publisher={APS}
}

@article{winnel2021overcoming,
  title={Overcoming the repeaterless bound in continuous-variable quantum communication without quantum memories},
  author={Winnel, Matthew S and Guanzon, Joshua J and Hosseinidehaj, Nedasadat and Ralph, Timothy C},
  journal={arXiv preprint arXiv:2105.03586},
  year={2021}
}

@article{erkilicc2023surpassing,
  title={Surpassing the repeaterless bound with a photon-number encoded measurement-device-independent quantum key distribution protocol},
  author={Erk{\i}l{\i}{\c{c}}, {\"O}zlem and Conlon, Lorc{\'a}n and Shajilal, Biveen and Kish, Sebastian and Tserkis, Spyros and Kim, Yong-Su and Lam, Ping Koy and Assad, Syed M},
  journal={npj Quantum Inf.},
  volume={9},
  number={1},
  pages={29},
  year={2023},
  publisher={Nature Publishing Group UK London}
}

@article{goebel2008multistage,
  title={Multistage entanglement swapping},
  author={Goebel, Alexander M and Wagenknecht, Claudia and Zhang, Qiang and Chen, Yu-Ao and Chen, Kai and Schmiedmayer, J{\"o}rg and Pan, Jian-Wei},
  journal={Phy. Rev. Lett.},
  volume={101},
  number={8},
  pages={080403},
  year={2008},
  publisher={APS}
}

@article{kaltenbaek2009high,
  title={High-fidelity entanglement swapping with fully independent sources},
  author={Kaltenbaek, Rainer and Prevedel, Robert and Aspelmeyer, Markus and Zeilinger, Anton},
  journal={Phys. Rev. A},
  volume={79},
  number={4},
  pages={040302},
  year={2009},
  publisher={APS}
}

@article{dur2007entanglement,
  title={Entanglement purification and quantum error correction},
  author={D{\"u}r, Wolfgang and Briegel, Hans J},
  journal={Rep. Prog. Phys.},
  volume={70},
  number={8},
  pages={1381},
  year={2007},
  publisher={IOP Publishing}
}

@article{verstraete2001local,
  title={Local filtering operations on two qubits},
  author={Verstraete, Frank and Dehaene, Jeroen and DeMoor, Bart},
  journal={Phys. Rev. A},
  volume={64},
  number={1},
  pages={010101},
  year={2001},
  publisher={APS}
}

@article{dur2003entanglement,
  title={Entanglement purification for quantum computation},
  author={D{\"u}r, Wolfgang and Briegel, H-J},
  journal={Phys. Rev. Lett.},
  volume={90},
  number={6},
  pages={067901},
  year={2003},
  publisher={APS}
}

@article{dehaene2003local,
  title={Local permutations of products of Bell states and entanglement distillation},
  author={Dehaene, Jeroen and Van den Nest, Maarten and De Moor, Bart and Verstraete, Frank},
  journal={Phys. Rev. A},
  volume={67},
  number={2},
  pages={022310},
  year={2003},
  publisher={APS}
}

@article{devetak2005distillation,
  title={Distillation of secret key and entanglement from quantum states},
  author={Devetak, Igor and Winter, Andreas},
  journal={Proc. R. Soc. A: Math. Phys. Eng. Sci.},
  volume={461},
  number={2053},
  pages={207--235},
  year={2005},
  publisher={The Royal Society}
}

@article{winnel2022achieving,
  title={Achieving the ultimate end-to-end rates of lossy quantum communication networks},
  author={Winnel, Matthew S and Guanzon, Joshua J and Hosseinidehaj, Nedasadat and Ralph, Timothy C},
  journal={npj Quantum Inf.},
  volume={8},
  number={1},
  pages={129},
  year={2022},
  publisher={Nature Publishing Group UK London}
}

@article{dias2020quantum,
  title={Quantum repeater for continuous-variable entanglement distribution},
  author={Dias, Josephine and Winnel, Matthew S and Hosseinidehaj, Nedasadat and Ralph, Timothy C},
  journal={Phys. Rev. A},
  volume={102},
  number={5},
  pages={052425},
  year={2020},
  publisher={APS}
}

@article{garcia2009reverse,
  title={Reverse coherent information},
  author={Garc{\'\i}a-Patr{\'o}n, Ra{\'u}l and Pirandola, Stefano and Lloyd, Seth and Shapiro, Jeffrey H},
  journal={Phys. Rev. Lett.},
  volume={102},
  number={21},
  pages={210501},
  year={2009},
  publisher={APS}
}

@article{shor1995scheme,
  title={Scheme for reducing decoherence in quantum computer memory},
  author={Shor, Peter W},
  journal={Phys. Rev. A},
  volume={52},
  number={4},
  pages={R2493},
  year={1995},
  publisher={APS}
}

@article{steane1996error,
  title={Error correcting codes in quantum theory},
  author={Steane, Andrew M},
  journal={Phys. Rev. Lett.},
  volume={77},
  number={5},
  pages={793},
  year={1996},
  publisher={APS}
}

@book{gottesman1997stabilizer,
  title={Stabilizer codes and quantum error correction},
  author={Gottesman, Daniel},
  year={1997},
  publisher={California Institute of Technology}
}

@article{calderbank1996good,
  title={Good quantum error-correcting codes exist},
  author={Calderbank, A Robert and Shor, Peter W},
  journal={Phys. Rev. A},
  volume={54},
  number={2},
  pages={1098},
  year={1996},
  publisher={APS}
}

@article{knill2001benchmarking,
  title={Benchmarking quantum computers: the five-qubit error correcting code},
  author={Knill, Emanuel and Laflamme, Raymond and Martinez, Rudy and Negrevergne, Camille},
  journal={Phys. Rev. Lett.},
  volume={86},
  number={25},
  pages={5811},
  year={2001},
  publisher={APS}
}

@article{wootton2018repetition,
  title={Repetition code of 15 qubits},
  author={Wootton, James R and Loss, Daniel},
  journal={Phys. Rev. A},
  volume={97},
  number={5},
  pages={052313},
  year={2018},
  publisher={APS}
}

@article{postol2001proposed,
  title={A proposed quantum low density parity check code},
  author={Postol, Michael S},
  journal={arXiv preprint quant-ph/0108131},
  year={2001}
}

@article{mackay2004sparse,
  title={Sparse-graph codes for quantum error correction},
  author={MacKay, David JC and Mitchison, Graeme and McFadden, Paul L},
  journal={IEEE Trans. Inf. Theory},
  volume={50},
  number={10},
  pages={2315--2330},
  year={2004},
  publisher={IEEE}
}

@article{kasai2011quantum,
  title={Quantum error correction beyond the bounded distance decoding limit},
  author={Kasai, Kenta and Hagiwara, Manabu and Imai, Hideki and Sakaniwa, Kohichi},
  journal={IEEE Trans. Inf. Theory},
  volume={58},
  number={2},
  pages={1223--1230},
  year={2011},
  publisher={IEEE}
}

@article{devitt2013quantum,
  title={Quantum error correction for beginners},
  author={Devitt, Simon J and Munro, William J and Nemoto, Kae},
  journal={Rep. Prog. Phys.},
  volume={76},
  number={7},
  pages={076001},
  year={2013},
  publisher={IOP Publishing}
}

@article{gisin2007quantum,
  title={Quantum communication},
  author={Gisin, Nicolas and Thew, Rob},
  journal={Nat. Photonics},
  volume={1},
  number={3},
  pages={165--171},
  year={2007},
  publisher={Nature Publishing Group UK London}
}

@article{pirandola2020advances,
  title={Advances in quantum cryptography},
  author={Pirandola, Stefano and Andersen, Ulrik L and Banchi, Leonardo and Berta, Mario and Bunandar, Darius and Colbeck, Roger and Englund, Dirk and Gehring, Tobias and Lupo, Cosmo and Ottaviani, Carlo and others},
  journal={Adv. Opt. Photonics},
  volume={12},
  number={4},
  pages={1012--1236},
  year={2020},
  publisher={Optica Publishing Group}
}

@article{xu2020secure,
  title={Secure quantum key distribution with realistic devices},
  author={Xu, Feihu and Ma, Xiongfeng and Zhang, Qiang and Lo, Hoi-Kwong and Pan, Jian-Wei},
  journal={Rev. Mod. Phys.},
  volume={92},
  number={2},
  pages={025002},
  year={2020},
  publisher={APS}
}

@article{dynes2009efficient,
  title={Efficient entanglement distribution over 200 kilometers},
  author={Dynes, James F and Takesue, Hiroki and Yuan, Zhiliang L and Sharpe, Andrew W and Harada, K and Honjo, Toshimori and Kamada, Hidehiko and Tadanaga, Osamu and Nishida, Yoshiki and Asobe, Masaki and others},
  journal={Opt. Express},
  volume={17},
  number={14},
  pages={11440--11449},
  year={2009},
  publisher={Optica Publishing Group}
}

@article{inagaki2013entanglement,
  title={Entanglement distribution over 300 km of fiber},
  author={Inagaki, Takahiro and Matsuda, Nobuyuki and Tadanaga, Osamu and Asobe, Masaki and Takesue, Hiroki},
  journal={Opt. Express},
  volume={21},
  number={20},
  pages={23241--23249},
  year={2013},
  publisher={Optica Publishing Group}
}

@article{wilde2017converse,
  title={Converse bounds for private communication over quantum channels},
  author={Wilde, Mark M and Tomamichel, Marco and Berta, Mario},
  journal={IEEE Trans. Inf. Theory},
  volume={63},
  number={3},
  pages={1792--1817},
  year={2017},
  publisher={IEEE}
}

@article{kimble2008quantum,
  title={The quantum internet},
  author={Kimble, H Jeff},
  journal={Nature},
  volume={453},
  number={7198},
  pages={1023--1030},
  year={2008},
  publisher={Nature Publishing Group}
}

@phdthesis{aschauer2005quantum,
  title={Quantum communication in noisy environments},
  author={Aschauer, Hans},
  year={2005},
  school={Ludwig–Maximilians–Universit\"{a}t}
}

@article{matsumoto2003conversion,
  title={Conversion of a general quantum stabilizer code to an entanglement distillation protocol},
  author={Matsumoto, Ryutaroh},
  journal={J. Phys. A Math. Gen.},
  volume={36},
  number={29},
  pages={8113},
  year={2003},
  publisher={IOP Publishing}
}

@article{ambainis2006minimum,
  title={The minimum distance problem for two-way entanglement purification},
  author={Ambainis, Andris and Gottesman, Daniel},
  journal={IEEE Trans. Inf. Theory},
  volume={52},
  number={2},
  pages={748--753},
  year={2006},
  publisher={IEEE}
}

@book{schuster2006deterministic,
  title={Deterministic chaos: an introduction},
  author={Schuster, Heinz Georg and Just, Wolfram},
  year={2006},
  publisher={John Wiley \& Sons}
}

@article{beaudry2011intuitive,
  title={An intuitive proof of the data processing inequality},
  author={Beaudry, Normand J and Renner, Renato},
  journal={arXiv preprint arXiv:1107.0740},
  year={2011}
}

@book{cover1999elements,
  title={Elements of information theory},
  author={Cover, Thomas M},
  year={1999},
  publisher={John Wiley \& Sons}
}

@article{beigi2013new,
  title={A new quantum data processing inequality},
  author={Beigi, Salman},
  journal={J. Math. Phys.},
  volume={54},
  number={8},
  year={2013},
  publisher={AIP Publishing}
}

@article{lindblad1975completely,
  title={Completely positive maps and entropy inequalities},
  author={Lindblad, G{\"o}ran},
  journal={Commun. Math. Phys.},
  volume={40},
  pages={147--151},
  year={1975},
  publisher={Springer}
}

@article{sheng2013hybrid,
  title={Hybrid entanglement purification for quantum repeaters},
  author={Sheng, Yu-Bo and Zhou, Lan and Long, Gui-Lu},
  journal={Phys. Rev. A},
  volume={88},
  number={2},
  pages={022302},
  year={2013},
  publisher={APS}
}

@article{zhou2020purification,
  title={Purification of the residual entanglement},
  author={Zhou, Lan and Zhong, Wei and Sheng, Yu-Bo},
  journal={Opt. Express},
  volume={28},
  number={2},
  pages={2291--2301},
  year={2020},
  publisher={Optica Publishing Group}
}

@article{huang2022experimental,
  title={Experimental one-step deterministic polarization entanglement purification},
  author={Huang, Cen-Xiao and Hu, Xiao-Min and Liu, Bi-Heng and Zhou, Lan and Sheng, Yu-Bo and Li, Chuan-Feng and Guo, Guang-Can},
  journal={Sci. Bull.},
  volume={67},
  number={6},
  pages={593--597},
  year={2022},
  publisher={Elsevier}
}

@article{hu2021long,
  title={Long-distance entanglement purification for quantum communication},
  author={Hu, Xiao-Min and Huang, Cen-Xiao and Sheng, Yu-Bo and Zhou, Lan and Liu, Bi-Heng and Guo, Yu and Zhang, Chao and Xing, Wen-Bo and Huang, Yun-Feng and Li, Chuan-Feng and others},
  journal={Phys. Rev. Lett.},
  volume={126},
  number={1},
  pages={010503},
  year={2021},
  publisher={APS}
}

@article{sheng2010deterministic,
  title={Deterministic entanglement purification and complete nonlocal Bell-state analysis with hyperentanglement},
  author={Sheng, Yu-Bo and Deng, Fu-Guo},
  journal={Phys. Rev. A},
  volume={81},
  number={3},
  pages={032307},
  year={2010},
  publisher={APS}
}

@article{sheng2010one,
  title={One-step deterministic polarization-entanglement purification using spatial entanglement},
  author={Sheng, Yu-Bo and Deng, Fu-Guo},
  journal={Phys. Rev. A},
  volume={82},
  number={4},
  pages={044305},
  year={2010},
  publisher={APS}
}

@article{riera2021entanglement,
  title={Entanglement-assisted entanglement purification},
  author={Riera-S{\`a}bat, Ferran and Sekatski, Pavel and Pirker, Alexander and D{\"u}r, Wolfgang},
  journal={Physical Review Letters},
  volume={127},
  number={4},
  pages={040502},
  year={2021},
  publisher={APS}
}

@article{shi2025stabilizer,
  title={Stabilizer Entanglement Distillation and Efficient Fault-Tolerant Encoders},
  author={Shi, Yu and Patil, Ashlesha and Guha, Saikat},
  journal={PRX Quantum},
  volume={6},
  number={1},
  pages={010339},
  year={2025},
  publisher={APS}
}

@article{bonilla2025constant,
  title={Constant-Overhead Fault-Tolerant Bell-Pair Distillation Using High-Rate Codes},
  author={Bonilla Ataides, J Pablo and Zhou, Hengyun and Xu, Qian and Baranes, Gefen and Li, Bikun and Lukin, Mikhail D and Jiang, Liang},
  journal={Phys. Rev. Lett.},
  volume={135},
  number={13},
  pages={130804},
  year={2025},
  publisher={APS}
}

@inproceedings{pattison2025constant,
  title={Constant-Rate Entanglement Distillation for Fast Quantum Interconnects},
  author={Pattison, Christopher and Baranes, Gefen and Bonilla Ataides, Juan Pablo and Lukin, Mikhail D and Zhou, Hengyun},
  booktitle={Proceedings of the 52nd Annual International Symposium on Computer Architecture},
  pages={257--270},
  year={2025}
}

\bibliographystyle{naturemag}

\end{document}